\documentclass[reprint, amsmath,amssymb, aps,pra]{revtex4-2}

\usepackage{graphicx}
\usepackage{dcolumn}
\usepackage{xcolor}
\usepackage{physics}
\usepackage[colorlinks,linkcolor=red,citecolor=blue,urlcolor=red]{hyperref}
\usepackage{float}
\usepackage[english]{babel}

\usepackage{comment}
\usepackage{mathrsfs}
\usepackage{amsmath}

\DeclareUnicodeCharacter{2009}{\,}

\renewcommand{\t}[1]{\mathrm{#1}}

\AtBeginDocument{%
    \newwrite\bibnotes
    \def\bibnotesext{Notes.bib}
    \immediate\openout\bibnotes=\jobname\bibnotesext
    \immediate\write\bibnotes{@CONTROL{REVTEX41Control}}
    \immediate\write\bibnotes{@CONTROL{%
    apsrev41Control,author="08",editor="1",pages="1",title="0",year="1"}}
     \if@filesw
     \immediate\write\@auxout{\string\citation{apsrev41Control}}%
    \fi
}%

\begin{document}

\preprint{APS/123-QED}

\title{Thermal intermodulation backaction in a high-cooperativity optomechanical system}

\author{Christian M. Pluchar}
\author{Aman R. Agrawal}
\author{Dalziel J. Wilson}
\affiliation{Wyant College of Optical Sciences, University of Arizona, Tucson, AZ 85721, USA}

\date{\today}
\begin{abstract}
The pursuit of room temperature quantum optomechanics with tethered nanomechanical resonators faces stringent challenges owing to extraneous mechanical degrees of freedom.
An important example is thermal intermodulation noise (TIN), a form of excess optical noise produced by mixing of thermal noise peaks. While TIN can be decoupled from the phase of the optical field, it remains indirectly coupled via radiation pressure, implying a hidden source of backaction that might overwhelm shot noise.  Here we report observation of TIN backaction in a high-cooperativity, room temperature cavity optomechanical system consisting of an acoustic-frequency Si$_3$N$_4$ trampoline coupled to a Fabry-P\'{e}rot cavity.  The backaction we observe exceeds thermal noise by 20 dB and radiation pressure shot noise by 40 dB, despite the thermal motion being 10 times smaller than the cavity linewidth. Our results suggest that mitigating TIN may be critical to reaching the quantum regime from room temperature in a variety of contemporary optomechanical systems.
\end{abstract}

\maketitle


Room temperature quantum experiments are a longstanding pursuit of cavity optomechanics~\cite{norte_mechanical_2016, purdy_quantum_2017, sudhir_quantum_2017}, spurred by the promise of fieldable quantum technologies~\cite{metcalfe2014applications, rademacher2020quantum, barzanjeh_optomechanics_2022} and simple platforms for fundamental physics tests \cite{gasbarri2021testing,carney2021mechanical}. Recently, ground state cooling has been achieved from room temperature using levitated nanoparticles \cite{delic2020cooling, magrini_real-time_2021}.  Ponderomotive squeezing has also been achieved at room temperature, using both levitated nanoparticles~\cite{magrini_squeezed_2022, militaru_ponderomotive_2022} and an optically stiffened cantilever \cite{aggarwal_room-temperature_2020}.  Despite this progress, however, including the recent development of ultracoherent nanomechanical resonators \cite{tsaturyan_ultracoherent_2017, ghadimi_elastic_2018}, room temperature quantum optomechanics with rigidly tethered nanomechanical resonators (e.g. strings and membranes) remains elusive, limited to signatures of weak optomechanical quantum correlations \cite{purdy_quantum_2017, sudhir_quantum_2017} and cooling to occupations of greater than $10$ \cite{guo2019feedback, saarinen2023laser, vezio2023optical}. Overcoming this hurdle is important because tethered nanomechanical resonators are readily functionalized and integrated with chip-scale electronics \cite{ekinci2005nanoelectromechanical}, features that form the basis for optomechanical quantum technologies~\cite{metcalfe2014applications, barzanjeh_optomechanics_2022}.

A key obstacle to room temperature quantum optomechanics is thermal intermodulation noise (TIN) \cite{fedorov_thermal_2020}, a form of excess optical noise produced in cavity optomechanical systems (COMS) due to the mixing of thermomechanical noise peaks.  TIN is especially pronounced in high-cooperativity tethered COMS \cite{leijssen2017nonlinear,fedorov_thermal_2020, beguin2020coupling, brawley2016nonlinear, guo2019feedback}, which commonly employ nanomechanical resonators with free spectral ranges orders of magnitude smaller than the cavity linewidth \cite{reinhardt_ultralow-noise_2016,tsaturyan_ultracoherent_2017}.  In conjunction with the cavity's transduction nonlinearity \cite{doolinNonlinearOptomechanicsStationary2014}, this high mode density can give rise to spectrally broadband TIN orders of magnitude in excess of shot noise \cite{fedorov_thermal_2020}---a severe impediment to
protocols that rely on the observability of quantum backaction, such as ground state cooling \cite{wilson_measurement-based_2015} and squeezing~\cite{purdy_strong_2013}. 

 \begin{figure}[ht!]
\centering  \includegraphics[width=1\columnwidth]{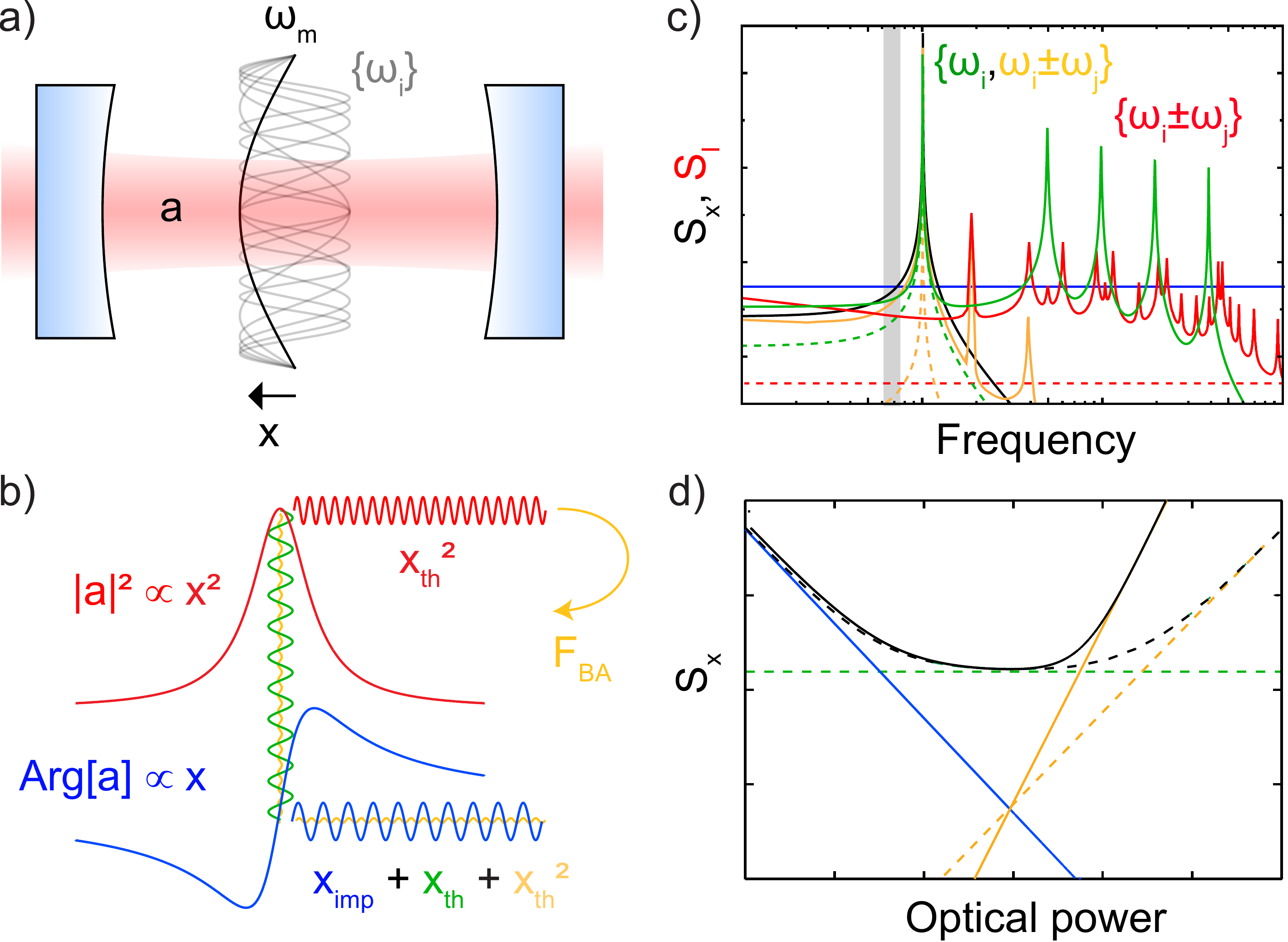}
    \caption{a) Multimode cavity optomechanical system (COMS) with intracavity field amplitude $a$ and mechanical resonance frequencies $\omega_i$. b) Nonlinear transduction of displacement $x$ into intensity (red) and linear transduction into phase (blue).  Radiation pressure (yellow) couples intensity and phase. c) Displacement and intensity ($I$) noise spectral density for resonantly probed COMS. Dashed lines: single mode COMS with intensity dominated by shot noise (red) and total displacement (black) by thermal noise (green) and QBA (orange).  Solid lines: multimode COMS with shot noise overwhelmed by TIN and QBA overwhelmed by TINBA.  d) Noise budget versus optical power for shaded region in (c).}
    \label{fig:Fig1}
\vspace{-5mm}
\end{figure}

Previous reports of TIN focus on its distortion of cavity-based measurement and methods to reduce it~\cite{fedorov_thermal_2020}.  These include reducing optomechanical coupling and cavity finesse, operating at a ``magic detuning" where the leading (quadratic) transduction nonlinearity vanishes \cite{fedorov_thermal_2020}, multi-mode cooling \cite{sommer2020multimode}, and enhancing mechanical $Q$. Proposals for cavity-free quantum optomechanics with ultracoherent nanomechanical resonators represent an extreme solution~\cite{pluchar_towards_2020,pratt_nanoscale_2021,hao_cavity-less_2020}.  A promising compromise exploits wide-bandgap phononic crystal nanoresonators in conjunction with broadband dynamical back-action cooling~\cite{saarinen2023laser}. Another key insight is that the phase of a resonant cavity probe is insensitive to TIN, allowing efficient feedback cooling even in the presence of thermal nonlinearity~\cite{fedorov_thermal_2020, saarinen2023laser}.

With this letter, we wish to highlight a complementary form of TIN---stochastic radiation pressure backaction (TINBA)---that is resilient to some of the above methods and poses an additional obstacle to room temperature quantum optomechanics. While previous studies have observed the effect of thermal nonlinearity on dynamical backaction---as an inhomogeneous broadening of the optical spring and damping effects \cite{leijssen2017nonlinear}---TINBA, a type of classical intensity noise heating,  is subtler, as it appears only indirectly in cavity-based measurements, and can dominate quantum backaction (QBA) even when the thermal nonlinearity is small. As a demonstration, we study TIN in a high cooperativity, room temperature COMS based on an acoustic-frequency Si$_3$N$_4$ trampoline coupled to Fabry-P\'{e}rot cavity (a popularly proposed system for room temperature quantum experiments \cite{norte_mechanical_2016}).  Despite the thermal motion of the trampoline being 10 times smaller than the cavity linewidth, we observe TINBA as high as 20 dB in excess of thermal noise and an estimated 40 dB in excess of QBA.  We show that this noise can be precisely modeled, despite its apparent complexity, and explore tradeoffs to mitigating it via its strong dependence on temperature, \mbox{finesse, and detuning.}

\section{Theory: QBA versus TINBA}

 As illustrated in Fig. \ref{fig:Fig1}, TINBA arises due to a combination of transduction nonlinearity and radiation pressure in cavity-based readout of a multimode mechanical resonator. We here consider the observability of QBA in the presence of TINBA, focusing on a single mechanical mode with displacement coordinate $x$ and frequency $\omega_\t{m}$.  As a figure of merit, we take the quantum cooperativity
\begin{equation}
\label{eq:Cq}
   C_q = \frac{S^\t{QBA}_x}{S_x^{\t{tot}} - S^\t{QBA}_x} \approx \frac{S^\t{QBA}_x}{S^\t{th}_x + S^\t{TIN}_x}
\end{equation}
where $S_x^\t{QBA}[\omega]$ is the single-sided power spectral density of displacement ($x$) produced by QBA, $S_x^{\t{tot}}[\omega]$ is the total displacement spectral density---including thermal motion $S_x^\t{th}[\omega]$ and TINBA $S_x^\t{TIN}[\omega]$, defined below---and $S_x[\omega_\t{m}]\equiv S_x$ denotes the spectral density on resonance.

To build a model for $C_q$, we first consider a single-mode COMS characterized by an optomechanical coupling
\begin{equation}
\omega_c(x) = \omega_\t{c,0}+G x.
\end{equation}
where $\omega_c$ is the cavity resonance frequency and $G$ is the optomechanical coupling strength.

In the small displacement limit, the coupled Langevin equations describing this system are \cite{aspelmeyer_cavity_2014}
\begin{subequations}\label{eq:COMEOM}\begin{align}
    &m\ddot{x}+m\Gamma \dot{x} + m\omega^2 x = F_{\t{th}}+\hbar G|a^2|\\
    &\dot{a} = (-i (\omega_0-\omega_c(x)) +\kappa)a + \sqrt{\kappa_\t{in}}s_\t{in}
\end{align}\end{subequations}
where Eq. \ref{eq:COMEOM}a describes the displacement of a mechanical oscillator with mass $m$, resonance frequency $\omega_\t{m}$, and damping rate $\Gamma_\t{m}$, driven by a thermal force $F_\t{th}$ and a radiation pressure backaction force $F_\t{BA} = \hbar G |a^2|$; and Eq. \ref{eq:COMEOM}b describes the complex amplitude $a$ of the cavity field with energy decay rate $\kappa$, driven at rate $\kappa_\t{in}$ by an input field with amplitude $s_\t{in}$ and frequency $\omega_0$, and normalized so that $|a|^2 = n_\t{c}$ is the intracavity photon number and $|s_\t{in}|^2$ is the input photon flux. 

Linearizing Eq. \ref{eq:COMEOM} about small fluctuations in $a$ yields 
\begin{equation}\label{eq:Sx}
S_x[\omega] = |\chi_\t{eff}(\omega)|^2 \left(S_F^\t{th}[\omega]+S_F^\t{QBA}[\omega]\right)
\end{equation}
where $\chi_\t{eff} (\omega) \approx (1/m)/(\omega^2 - \omega_\t{eff}^2 - i \Gamma_\t{eff} \omega)$ is the effective mechanical susceptibility accounting for optical stiffening $k_\t{opt} = m\omega_\t{eff}^2-m\omega_\t{m}^2$ and damping $\Gamma_\t{opt} = \Gamma_\t{eff}-\Gamma_\t{m}$ \cite{aspelmeyer_cavity_2014},
\begin{equation}
    \label{eq:SFth}
    S^\t{th}_F[\omega] \approx 4  k_B T m \Gamma_m, 
\end{equation}
is the thermal force power spectral density \cite{saulson_thermal_1990} assuming a bath temperature $T\gg \hbar\omega_m/k_B$, and 
\begin{equation}
    \label{eq:SFQBA}
    S_{F}^\t{QBA} [\omega]= (\hbar G n_\t{c})^2 S_\t{RIN}^\t{shot}[\omega] = (\hbar G n_c)^2\frac{8/(n_\t{c}\kappa)}{1+4(\frac{\omega+\Delta}{\kappa})^2}
\end{equation}
is the QBA force produced by shot fluctuations of the intracavity photon number $S_{n_\t{c}}^\t{shot}[\omega]$ \cite{marquardt2007quantum,aspelmeyer_cavity_2014}, here expressed as a relative intensity noise $S_\t{RIN}^\t{shot}[\omega]= S_{n_\t{c}}^\t{shot}[\omega]/n_\t{c}^2$, and $\Delta = \omega_0 - \omega_{c,0}$ is the laser-cavity detuning.

We hereafter specialize to the fast cavity limit $\omega_m\ll\kappa$, in which most 
room temperature quantum optomechanics experiments with tethered mechanical resonators operate, including ours.  In this case $C_q > 1$ requires
\begin{equation}
    \label{eq:QBAoverTH}
   \frac{S_F^\t{QBA}}{S_F^\t{th}} = \frac{C_0 n_c}{n_\t{th}}\frac{1}{1+4\Delta^2/\kappa^2}>1
\end{equation}
where $n_\t{th} = \frac{k_B T}{\hbar\omega_0}$ is the thermal bath occupation and 
\begin{equation}\label{eq:C0}
    C_0 = \frac{ 2G^2 \hbar }{m\omega_m\Gamma_m\kappa} = \frac{4 g_0^2}{\kappa\Gamma_m}
\end{equation}
is the vacuum optomechanical cooperativity, expressed on the right-hand side in terms of the vacuum optomechanical coupling rate $g_0 = Gx_\t{ZP}$, where $x_\t{ZP}= \sqrt{\hbar/(2m\omega_m)}$ is the oscillator's zero-point motion.

Turning to TIN, we first re-emphasize that Eq. \ref{eq:COMEOM} considers only a single mechanical mode whose displacement is perturbatively small, $G x \ll \kappa$.   In practice, however, operating in the fast cavity limit using tethered nanomechanical resonators usually implies that many mechanical modes are simultaneously coupled to the cavity:
\begin{equation}
    \omega_\t{c} = \omega_\t{c,0}+\sum_i G_i x_i.
\end{equation}
Moreover, for a single mode designed for high cooperativity (Eq. \ref{eq:C0}), low stiffness $m\omega_\t{m}^2$ and high temperature can readily lead to a root-mean-squared thermal displacement $x_\t{th}$ exceeding the nonlinear transduction threshold   
\begin{equation}\label{eq:nonlinearcondition}
x_\t{th} = \sqrt{\frac{k_B T}{m\omega_m^2}} \gtrsim \frac{\kappa}{G} \sim \frac{\lambda}{\mathcal{F}}
\end{equation}
where $\mathcal{F}$ is the cavity finesse.  As explored in \cite{fedorov_thermal_2020}, the combination of these features---multiple mechanical modes exhibiting thermal nonlinearity---can lead to broadband TIN $S_{n_\t{c}}^\t{TIN}$ due to the mixing of thermal noise peaks.  This in turn gives rise to a TINBA force
\begin{equation}\label{eq:TINBA_1}
S_F^\t{TIN}[\omega]=(\hbar G n_\t{c})^2 S_\t{RIN}^\t{TIN}[\omega]
\end{equation}
which, like classical laser intensity noise \cite{verlotExperimentalDemonstrationQuantum2011}, can drive the mechanical resonator in excess of QBA.

To analyze TINBA in the fast cavity limit, it suffices to consider the steady-state dependence of $n_\t{c}$ on detuning, expanded to second order in deviations from the mean value $\Delta$.  For convenience, following \cite{fedorov_thermal_2020}, we define the relative detuning $\nu = 2\Delta /\kappa$ and its deviation $\delta \nu$, yielding
\begin{equation}\label{eq:nc_expansion}
n_\t{c}(\delta\nu) \propto 1-\frac{2\nu}{1+\nu^2}\delta\nu+\frac{3\nu^2-1}{(1+\nu^2)^2}\delta \nu^2.
\end{equation}
For a single mechanical mode, with $\delta \nu = 2G x/\kappa$, the second term in Eq. \ref{eq:nc_expansion} corresponds to the optical spring force $F_\t{BA}(x)= k_\t{opt}(\nu) x$.  For a multimode optomechanical system, with $\delta\nu = \sum_n 2 G_n x_n/\kappa$, the third term gives rise to intermodulation noise.  To see this, using \textcolor{black}{
Wick's theorem \cite{isserlisFormulaProductMomentCoefficient1918}}, the spectrum of $\nu^2$ can be expressed as the self-convolution of double-sided linear spectrum $S_{\nu\nu}[\omega]$
\begin{equation}
\label{eq:Snu2}
    S_{\nu^2}[\omega] =  4 \int^{\infty}_{-\infty} S_{\nu \nu} [\omega'] S_{\nu\nu} [\omega - \omega'] \frac{d\omega'}{2\pi}, 
\end{equation}
where 
\begin{equation}
    \label{eq:Snunu}
S_{\nu \nu}[\omega] = \sum_n \frac{4 G_n^2}{\kappa^2} S_{xx}^n[\omega]
\end{equation}
is the cavity frequency noise including all mechanical modes for which $\omega_n\lesssim\kappa$. The resulting TIN
\begin{equation}\label{eq:SncTIN}
S_\t{RIN}^\t{TIN}[\omega] = \frac{(3\nu^2-1)^2}{(1+\nu^2)^4} S_{\nu^2}[\omega]
\end{equation}
gives rise to a TINBA force (Eq. \ref{eq:TINBA_1})
\begin{equation}\label{eq:SFTIN}
S_F^\t{TIN}[\omega] = (\hbar G n_\t{c})^2 \frac{(3\nu^2-1)^2}{(1+\nu^2)^4} S_{\nu^2}[\omega].
\end{equation}

Three features of TINBA bear emphasis.  First, unlike QBA or photothermal heating (which both scale as $S_x\propto n_c$), TINBA scales quadratically with $n_c$.  Second, unlike the optical spring, TINBA does not vanish on resonance ($\nu = 0$).  In fact, it is maximal in this case, simplifying to\begin{equation}
    \label{eq:TINBA}
   S_F^\t{TIN}[\omega,\nu = 0] = (\hbar G n_\t{c})^2 S_{\nu^2}[\omega]
\end{equation}
corresponding to $S_\t{RIN}^\t{TIN}[\omega] = S_{\nu^2}[\omega]$.  Third, there exists a ``magic'' detuning $|\nu |= 1/\sqrt{3}$ at which TINBA vanishes, 
\begin{equation}
   S_F^\t{TIN}[\omega,\nu = \pm 1/\sqrt{3}] = 0
\end{equation}
corresponding to a $\partial^2 n_c/\partial^2 \nu = 0$.  However, at this detuning, the optical spring is maximized, possibly leading to instability ($k_\t{opt}\approx -m\omega_m^2$) for large $n_\t{c}$.

\begin{figure*}[t!]
\vspace{-3mm}

    \centering
    \includegraphics[width=2\columnwidth]{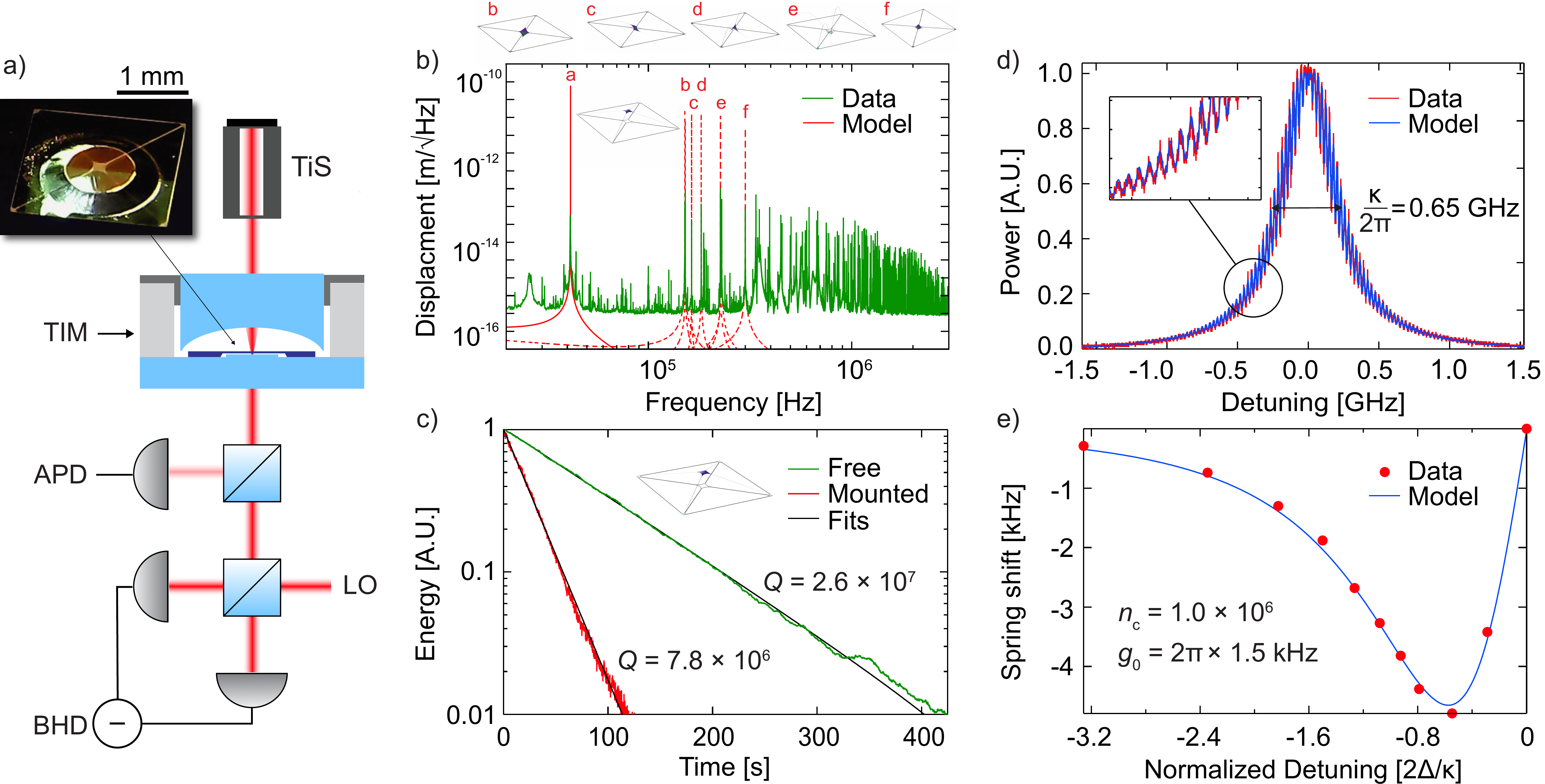}
    \caption{Trampoline-in-the-middle (TIM) cavity optomechanical system. a) Experimental setup: the TIM cavity is driven by a Titanium Sapphire laser (TiS) and probed in transmission with an avalanche photodiode (APD) and balanced homodyne detector (BHD). The BHD local oscillator (LO) is derived from the same TiS.  b) Displacement measurement using a resonant probe and the BHD tuned to the phase quadrature.  Thermal noise models for the first five trampoline modes (a-d) are overlayed. c) Ringdown measurements of the fundamental trampoline before (green) and after (red) mounting inside the TIM cavity. d) Detuning sweep over cavity resonance, fit with a model that includes the thermal motion of the trampoline fundamental mode. e) Measurement of the optical spring shift versus detuning, used to determine the intracavity photon number.}
    \label{fig:Fig2}
\vspace{-3mm}
\end{figure*}

With these features in mind, we suggest three conditions for observing QBA ($C_q\gtrsim 1$) in the presence of TIN, valid for all detunings in the bad cavity limit:
\begin{subequations}\label{eq:QBAconditions}\begin{align}
    n_\t{c} &\gtrsim (1+\nu^2)\frac{n_\t{th}}{C_0}\\
    Q_\t{m} &\gtrsim \frac{2\nu}{1+\nu^2}n_\t{th}\\
    S_\t{RIN}^\t{TIN}&
    \lesssim \frac{1}{(1+\nu^2)^2}\frac{2 S_\nu^\t{ZP}}{n_\t{th}}
\end{align}\end{subequations}
where $Q_\t{m}=\omega_\t{m}/\Gamma_\t{m}$ is the mechanical quality factor and $S_\nu^\t{ZP} = (4g_0^2/\gamma)/ \kappa^2$ is the normalized zero-point detuning spectral density, related to the zero-point displacement spectral density $S_x^\t{ZP} =  S_x^\t{th}/(2n_\t{th}) = 4 x_\t{ZP}^2/\Gamma_m$ by $S_\nu^\t{ZP}=G^2 S_x^\t{ZP}/\kappa^2$.  The first two conditions are independent of TIN and correspond to $S_F^\t{QBA}>S_F^\t{th}$ and $\omega_\t{eff}>0$, respectively.  The last condition implies $S_F^\t{TIN}<S_F^\t{th}$ when $n_\t{c}>n_\t{th}/C_0$, and is given by minimizing the relation
\begin{equation}\label{eq:CqwithTIN}
C_q=\frac{1}{1+\nu^2}\left(\frac{S_\t{RIN}^\t{TIN}\textcolor{black}{(G,\kappa,\nu)}}{8/\kappa}n_c+\frac{n_\t{th}}{C_0 n_\t{c}}\right)^{-1}
\end{equation}
where we have emphasized the depedence of $S_\t{RIN}^\t{TIN}$ on system parameters and detuning.

In the next section, we explore the requirements in Eq. \ref{eq:QBAconditions} in a popular membrane-in-the-middle platform, and show \ref{eq:QBAconditions}c may not be met even if \ref{eq:QBAconditions}a and \ref{eq:QBAconditions}b are.

\section{Trampoline-in-the-middle system}

Our optomechanical system consists of a Si$_3$N$_4$ trampoline resonator coupled to a Fabry-P\'{e}rot cavity in the membrane-in-the-middle configuration \cite{thompson_strong_2008}---hereafter dubbed ``trampoline-in-the-middle'' (TIM).  
As shown in Fig. \ref{fig:Fig2}a, the quasi-monolithic cavity is assembled by sandwiching the Si device chip between a concave (radius 10 cm) and a plano mirror, with the trampoline positioned nearer the plano mirror to minimize diffraction loss.  (This is achieved by etching the plano mirror into a mesa-like structure \cite{noauthor_supplementary_nodate}, as shown in Fig. \ref{fig:Fig2}a.)  A relatively short cavity length of $L = 415\;\mu\t{m}$ is chosen, to reduce sensitivity to laser frequency noise \cite{noauthor_supplementary_nodate}.
The bare (without trampoline) cavity finesse is as high as $\mathcal{F}=3\times 10^4$ at wavelengths near the coating center, $\lambda = 850$ nm. To reduce thermal nonlinearity, $\mathcal{F}$ is reduced to $\approx 550$ by operating at $\lambda\approx 786$ nm, yielding a cavity decay rate of $\kappa = 2\pi\times 0.65$ GHz.

\begin{figure}
\vspace{-3mm}
    \centering
    \includegraphics[width=1\columnwidth]{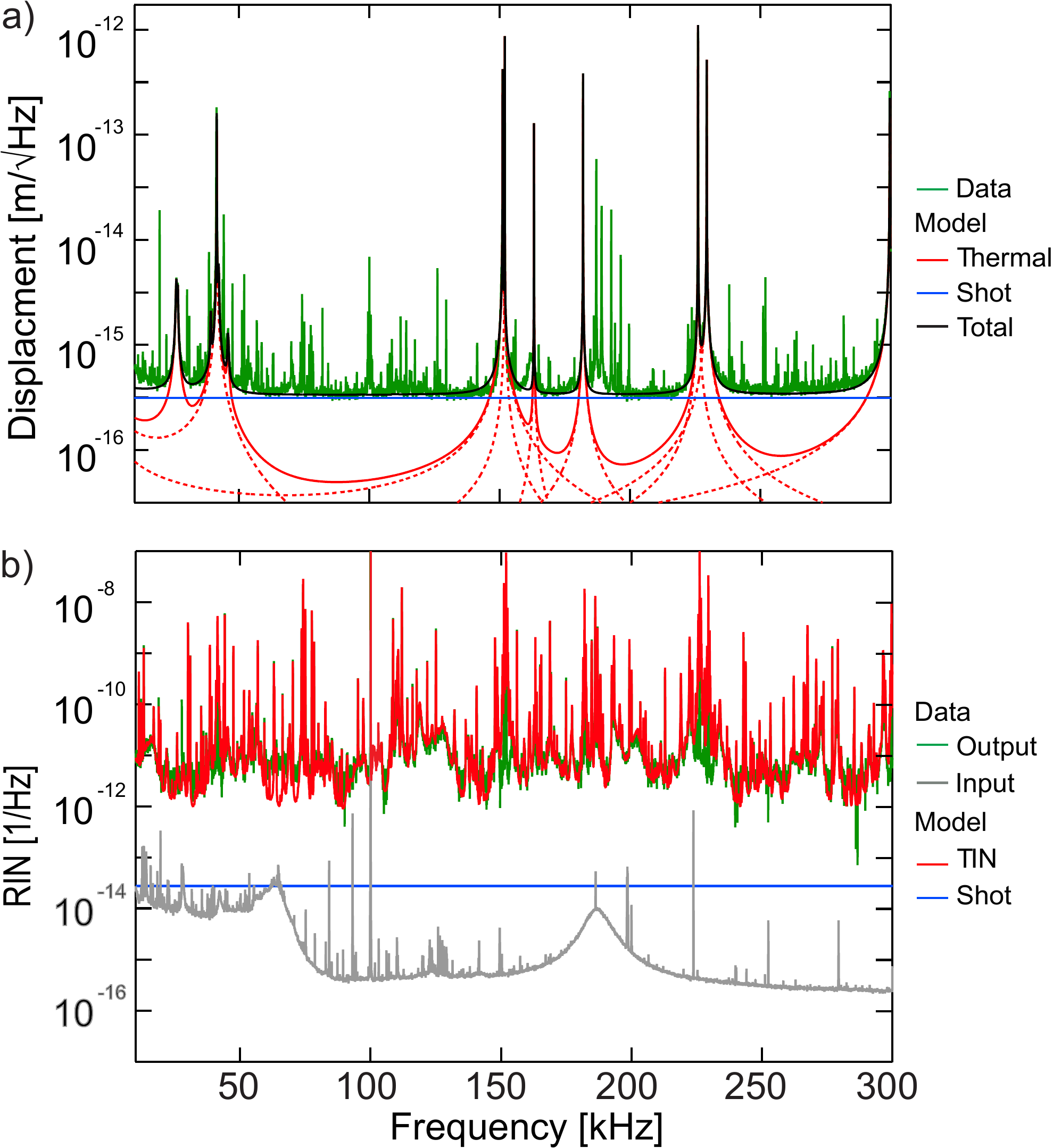}
    \caption{Thermal intermodulation noise (TIN) in TIM system. a) Low-frequency part of the phase measurement in Fig. \ref{fig:Fig2}b (green), compared to models of the thermal motion of the first seven trampoline modes (dotted red), their incoherent sum (solid red), shot noise (blue), and total noise (black). b) Simultaneously recorded intensity noise (green), normalized by the mean intensity (RIN).  Red curve is a TIN model (see main text), blue is a shot noise model, and gray is the measurement of the input laser RIN, after the noise eater \cite{noauthor_supplementary_nodate}.}
    \label{fig:Fig3}
\vspace{-5mm}
\end{figure}

\begin{figure*}[t!]
\vspace{-3mm}
    \centering
     \includegraphics[width=1.7\columnwidth]{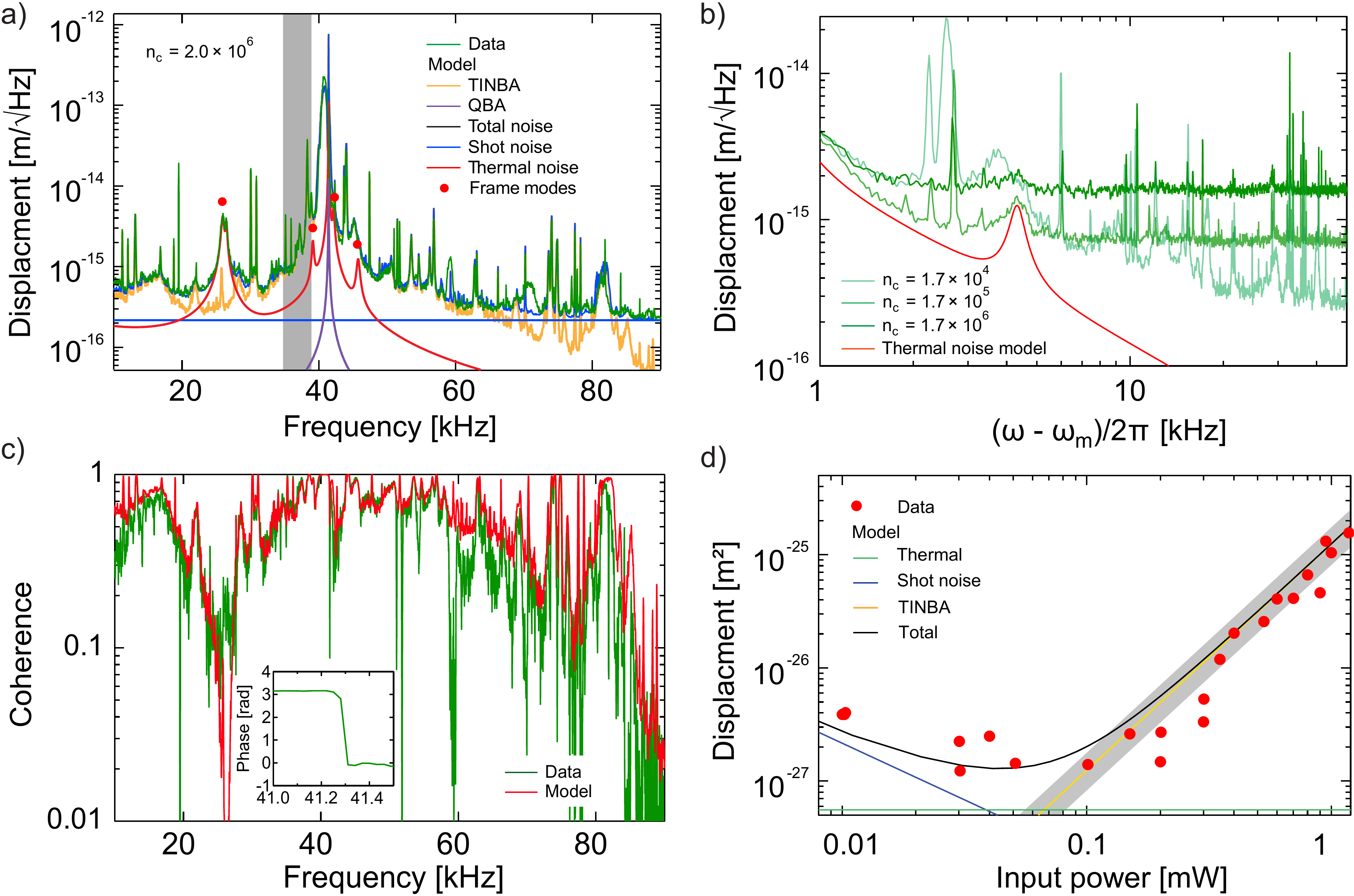}
    \caption{ Evidence for TINBA in a strong displacement measurement.  a) Displacement noise near the fundamental trampoline resonance $\omega_\t{m}$, with models for various components overlaid and shading for integration region in (d). b) Noise at different powers versus offset frequency $\omega-\omega_m$. Shot noise drops inversely with power at large offsets.  Near resonance, TINBA-driven motion increases quadratically with power. c) Coherence between the noise in (a) and a simultaneously recorded intensity noise. Values near unity imply a high degree of correlation. Inset: phase of the coherence function.  The $\pi$ phase change on resonance is due the mechanical susceptibility. d)  Noise integrated over the frequency range shaded in (a), overlaid with models for shot noise (blue), thermal noise (green), and TINBA (yellow). Gray shading is the uncertainty of the TINBA model due to statistical variation of the measured TIN. }
    \label{fig:Fig4}
\vspace{-4mm}
\end{figure*}

Si$_3$N$_4$ trampolines are popular candidates for room temperature quantum optomechanics experiments \cite{reinhardt_ultralow-noise_2016,norte_mechanical_2016, pluchar_towards_2020}, owing to their high $Q$-stiffness ratio.  We employ an 85-nm-thick trampoline with a $200\times 200\,\mu\t{m}^2$ central pad, $1700\times4.2\,\mu\t{m}^2$ tethers, and tether fillets designed to optimize strain-induced dissipation dilution of the fundamental mode \cite{sadeghi2019influence}, yielding $Q_\t{m} = 2.6\times 10^7$ \cite{pluchar_towards_2020}, $\omega_\t{m} \approx 2\pi\times 41$ kHz, and a COMSOL-simulated effective mass of $m = 12$ ng. 
Care was taken to mount the trampoline without introducing loss; nevertheless, two small dabs of adhesive reduced $Q_\t{m}$ to $7.8\times 10^6$ (we speculate that this is due to hybridization with low-Q chip modes \cite{de_jong_mechanical_2022}, as evidenced by the satellite noise peaks in Fig. \ref{fig:Fig4}.)  The resulting thermal force noise $S_F^\t{th} \approx 8 \times 10^{-17}\,\t{N}/\sqrt{\t{Hz}}$ is in principle sufficient to observe QBA with an input power of $\sim 10\,\t{mW}$ at $\mathcal{F}\sim 10^3$.  Challenging this prospect is the fact that the trampoline's thermal motion, $x_\t{th} = 0.07$ nm, near the nonlinear transduction threshold (Eq. \ref{eq:nonlinearcondition}) at $\mathcal{F}\sim 10^3$.  Moreover, $\kappa/\omega_m\sim 10^4$ allows many higher-order trampoline modes to be coupled to the cavity field (Fig. 2b), satisfying the conditions for strong TIN.  

Measurements characterizing the nonlinearity and cooperativity of our TIM system are shown in Fig. \ref{fig:Fig2}d-e. A hallmark of thermal nonlinearity is modulation of the steady-state cavity response, as shown in Fig. \ref{fig:Fig2}d. Here the cavity length is swept across resonance with strong optomechanical coupling, corresponding to the trampoline positioned between nodes of the intracavity standing wave \cite{jayich_dispersive_2008}.  Fitting the transmitted power to $P_\t{out}(\nu(t)) \propto 1/(1+(\nu +8G x_\t{th} \cos(\omega_\t{m}\nu/\dot{\nu})/\kappa))^2)$ yields $G x_\t{th}/\kappa\approx 0.04$ \cite{noauthor_supplementary_nodate}, corresponding to $g_0 = G x_\t{th}/\sqrt{2 n_\t{th}}\sim 2\pi\times 1\,\t{kHz}$.
A more careful estimate of $g_0 = 2\pi\times 1.5\,\t{kHz}$ was obtained using the frequency sideband calibration method \cite{gorodetksy_determination_2010}, suggesting a vacuum cooperativity of $C_0 \approx 3$.  In the experiment below, for each input power $P_\t{in}$, the intracavity photon number $n_\t{c}$ is determined by recording the optical spring shift versus detuning and comparing to a model, $\Delta \omega_\t{opt}\approx k_\t{opt}/(2m\omega_m) \approx \Gamma_\t{m}C_0 n_\t{c}(\nu=0) \nu/(1+\nu^2)^2 $ (Fig. \ref{fig:Fig2}e).  For all measurements, the thermal nonlinearity is reduced by measurement-based feedback cooling of the fundamental mode. Full details and methods are presented in the SI \cite{noauthor_supplementary_nodate}.

\section{Observation of TIN backaction}

We now explore TIN in our TIM system and present evidence that TINBA overwhelms QBA at a sufficiently high intracavity photon number.  To this end, the cavity is probed on resonance ($\nu = 0$) with a Titanium-Sapphire laser (M-Squared Solstis) prestabilized with a broadband electro-optic intensity noise eater \cite{noauthor_supplementary_nodate}.  The output field $s_\t{out} \propto \sqrt{\kappa}a$ is monitored with two detectors.  A balanced homodyne detector records the phase of $s_\t{out}$, which encodes the trampoline displacement,  $\t{Arg}[s_\t{out}]\propto x$ with a shot-noise-limited imprecision of $S_x^\t{imp} \geq S_x^\t{ZP}/(8 C_0 n_c)$.  An avalanche photodiode (APD) records the output intensity $|s_\t{out}|^2 \propto n_\t{c}$ and is also used to lock the cavity using the Pound-Drever-Hall technique~\cite{noauthor_supplementary_nodate}.  

TIN couples directly to intracavity intensity and indirectly to mechanical displacement, via TINBA.  We explore this in Fig. \ref{fig:Fig3}, by comparing the intensity and phase fluctuations of an optical field passed resonantly through the TIM cavity.  An input power of $P_\t{in} = 0.2\;\t{mW}$ is used, corresponding to $n_c = 4\eta P_\t{in}/(\hbar\omega_c\kappa) = 4\times 10^5$ intracavity photons (with $\eta=0.4$ determined from the optical spring shift) 
and an ideal quantum cooperativity of $C_q = 7 \times 10^{-3} $.  The phase noise spectrum is calibrated in displacement units by bootstrapping to a model for the fundamental thermomechanical noise peak \cite{noauthor_supplementary_nodate}, yielding the apparent displacement noise spectrum,
\begin{equation}
    S_y[\omega]\equiv S_x[\omega]+S_x^\t{imp}[\omega].
\end{equation}
As seen in Fig. \ref{fig:Fig3}a, $S_y[\omega]$ is dominated by thermal noise near mechanical resonances and shot noise far from resonance.  The intensity noise (Fig. \ref{fig:Fig3}b) meanwhile exceeds shot noise and features numerous peaks at intermediate frequencies, 
suggestive of TIN.  To confirm this hypothesis, in \ref{fig:Fig3}b, we overlay the measured intensity noise with our TIN model (Eq. \ref{eq:Snu2}) using the phase-noise-inferred thermomechanical noise $S_y[\omega]-S_x^\t{imp}[\omega] \approx S_x^\t{th} [\omega]$ as an input. We observe strong agreement over the full measurement band, as highlighted in Fig. \ref{fig:Fig3}b. 

We now turn our attention to the spurious peaks in the phase noise spectrum in Fig. \ref{fig:Fig3}a, which we argue is displacement produced by TINBA.  To this end, in Fig.~\ref{fig:Fig4}, we compare $S_y[\omega]$ at different probe powers, focusing on Fourier frequencies near the fundamental trampoline resonance. Qualitatively, as shown in Fig. \ref{fig:Fig4}a-b, 
we observe an increase in the apparent displacement at larger powers, with a shape that is consistent with the measured intensity noise multiplied by the mechanical susceptibility.  To confirm that this is TINBA, in Fig. \ref{fig:Fig4}d we plot $S_y[\omega_m+\delta]$ at an offset $\delta \gg \Gamma_\t{m}$ versus versus input power in the range $P_\t{in}\in[10\,\mu\t{W},1.1\,\t{mW}]$ ($n_\t{c}\in [1.7\times 10^4,2.2\times 10^6]$). The observed quadratic scaling with $P_\t{in}$ is consistent with TINBA and distinct from QBA and photoabsorption heating.  The absolute magnitude of the displacement moreover agrees quantitatively well with our TINBA model, $S_x[\omega_m+\delta]  \approx S_\t{RIN}[\omega_m+\delta]/(m\omega_m^2\gamma_m^2)^2$ (black line), allowing for statistical uncertainty (gray shading) due to fluctuations in the measured $S_\t{RIN}[\omega]$. 

As an additional consistency check, we measure the coherence between the phase and intensity noise, allowing us to rule out artifacts such as imperfect cancellation of intensity noise in the balanced homodyne receiver. 
We define the coherence between signals $a$ and $b$ as $C_{ab}[\omega] = |S_{ab}[\omega]|^2 / (S_a[\omega] S_b[\omega])$ \cite{purdy_observation_2013}, where $S_{ab}[\omega]$ is the cross-spectrum of $a$ and $b$, so that $C_{ab}[\omega] = 1$ for perfectly correlated or anti-correlated signals, and $\t{Arg}[S_{ab}]$ characterizes the relative phase of the correlated signal components.  Fig. \ref{fig:Fig4}c shows the coherence of the phase and intensity noise measurements with $n_c = 2 \times 10^6$, together with a model that predicts the coherence based on the measured TIN noise and the mechanical susceptibility.  A high degree of coherence is observed over a $\sim 100$ kHz bandwidth surrounding the fundamental resonance.  Moreover, near resonance, the argument of the coherence undergoes a $\pi$-phase shift, indicative of the response of the mechanical susceptibility, and consistent with TINBA-driven motion.
Full details about models and measurements can be found in the SI~\cite{noauthor_supplementary_nodate}.

\section{Implications for quantum optomechanics experiments}

TIN backaction currently limits the quantum cooperativity of our TIM system.  For example, at the highest intracavity photon number in Fig. \ref{fig:Fig4}, $n_c = 2\times 10^6$,  $S_x^\t{TIN}/S_x^\t{th}\approx 10^2$ and $S_x^\t{TIN}/S_x^\t{QBA} = (S_x^\t{TIN}/S_x^\t{th})/C_q^0  \approx 2500$, implying that $C_q\approx 4\times 10^{-4}$ instead of the ideal value, $C_q^0 = C_0 n_\t{c}/n_\t{th} = 0.04$.  As implied by Eq. \ref{eq:QBAconditions}c, this could have been anticipated by comparing the measured TIN to the a priori zero-point frequency noise $S_\nu^\t{ZP} = C_0/\kappa\approx 7\times 10^{-10}$ scaled by the thermal occupation $n_\t{th}\approx 1.5\times 10^8$.  In our case, $S_\t{RIN}^\t{TIN}\approx 10^{-11}$ Hz$^{-1}$, yielding $C_q\lesssim 10^{-3}$ according to Eq.~\ref{eq:CqwithTIN}:
\begin{equation}\label{eq:Cqmin_resonant}
C_q(\nu = 0)=\left(\frac{S_\t{TIN}^\t{RIN}}{8/\kappa}n_c+\frac{n_\t{th}}{C_0 n_\t{c}}\right)^{-1}\le \sqrt{\frac{2S_\nu^\t{ZP}/n_\t{th}}{S_\t{TIN}^\t{RIN}}}
\end{equation}
The lower bound of Eq. \ref{eq:Cqmin_resonant} (and more generally, Eq. \ref{eq:CqwithTIN}) applies to any form of classical intensity noise, and results from the fact that classical intensity noise increases quadratically faster with $n_\t{c}$ than shot noise.  There is, therefore, always a probe strength at which classical backaction overwhelms QBA. To increase this threshold, one must either increase the zero-point spectral density  or reduce the intensity noise, leveraging, if possible, the different scaling of these noise terms with system parameters.  

\begin{figure*}[t!]
    \centering
    \includegraphics[width=2\columnwidth]{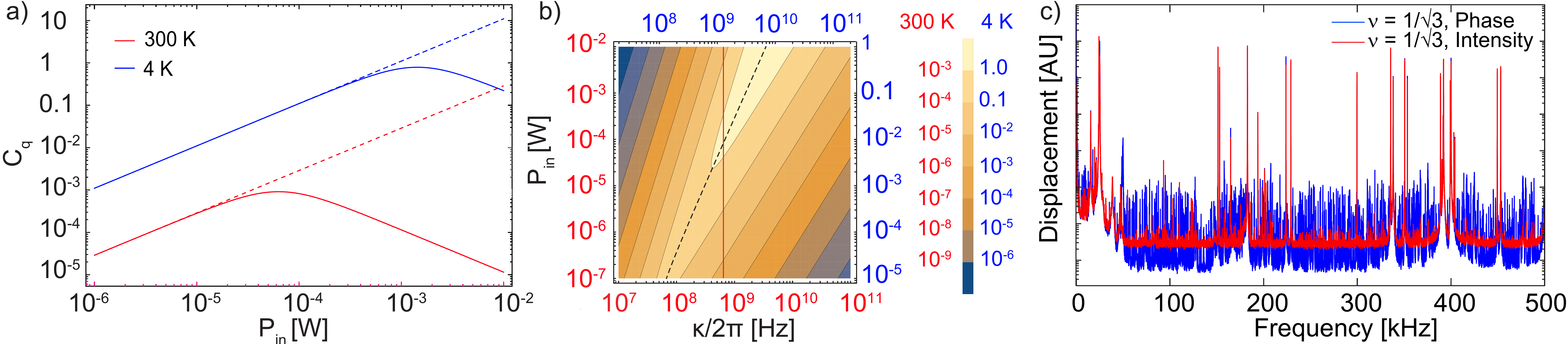}
    \caption{Strategies for mitigating TINBA.  a) Estimated quantum cooperativity $C_q$ of TIM system at $T = 300$ K (red) and 4 K (blue) assuming no TIN (dashed line) and parameters in the experiment (solid line, $S^{\t{TIN}}_{\t{RIN}} \approx 10^{-11}$, $g_0 = 2 \pi \times 1.5$ kHz and $\kappa = 2 \pi \times 650$ MHz).  Cryogenic operation reduces thermal noise and TIN, increasing the maximum $C_q$ by $1/T^2$. b) $C_q$ of TIM system as a function of linewidth and input power, for $T = 300$ K and 4 K. The red line indicates the slice shown in (a).  The dashed line indicates the maximum $C_q$ for a given $\kappa$ and $P_{\t{in}}$. c) Readout of the TIM cavity at the ``magic detuning,'' showcasing a decrease in TIN in the intensity noise (red), but an increase of TIN in the phase (displacement) noise (blue).}
    \label{fig:Fig5}
\end{figure*}

In the case of TINBA, which originates from thermal nonlinearity (Eq. \ref{eq:nonlinearcondition}), inroads can be made by leveraging the dependence of the nonlinearity on $G$, $\kappa$, and $\nu$.  
For example, the fact that $S_{F}^\t{TIN}\propto (G/\kappa)^4$ and $S_{F}^\t{QOBA}\propto (G/\kappa)^2$ suggests that TINBA can be mitigated by using a higher $\kappa$ (lower finesse $\mathcal{F}$) cavity.  In Fig. \ref{fig:Fig5}b we consider this strategy for our TIM system using the above experimental parameters and $\nu=0$.  Evidently, $C_q\sim 1$ is possible with 100-fold lower $\mathcal{F}$; however, it would require a proportionately larger laser power. This is problematic because of photothermal heating and increased demands on classical laser intensity noise suppression.  Fig. \ref{fig:Fig5}a-b shows the same computation at $T = 4$ K, revealing that power demands are relaxed in proportional to $T$, since $S_F^\t{TIN}\propto T^2$. This observation re-affirms the demands of room temperature quantum optomechanics and, conversely, the advantages of cryogenic pre-cooling.

Finally, we re-emphasize the strong detuning dependence of TIN and TINBA.  Operating at the magic-detuning $\nu = 1/\sqrt{3}$, as shown in Fig. \ref{fig:Fig5}c, can eliminate TIN in the optical  intensity; however, as evident in the blue data, the phase response of the cavity becomes nonlinear, potentially preventing the observation of quantum correlations generated via the optomechanical interaction. Moreover, in the regime of strong QBA, the associated optical spring (maximal at $\nu = 1/\sqrt{3}$ in the fast cavity limit) can be substantial.  To avoid instability ($\omega_\t{eff}=0$), one strategy is to use a dual-wavelength probe with $\nu = \pm 1/\sqrt{3}$, but this doesn't resolve the phase nonlinearity issue.  Another promising strategy---not considered in our theoretical analysis---is to exploit optical damping at $\nu\ne 0$ to realize multi-mode cooling \cite{saarinen2023laser}. The success of this strategy will
depend on the details of the system and may benefit by operating in an intermediate regime between the fast and slow cavity limit.  

\section{Conclusion}
We have explored the effect of thermal intermodulation noise (TIN) on the observability of radiation pressure quantum backaction (QBA) in a room temperature cavity optomechanical system and argued that TIN back-action  (TINBA) can overwhelm QBA under realistic conditions.  As an illustration, we studied the effects of TINBA in a high-cooperativity trampoline-in-the-middle cavity optomechanical system and found that it overwhelmed thermal noise and QBA by several orders of magnitude.  The conditions embodied by our TIM system---transduction nonlinearity, large thermal motion, and a multi-mode mechanical resonator---can be found in a wide variety of room-temperature quantum optomechanics experiments based on tethered nanomechanical resonators, including an emerging class of systems based on ultracoherent phononic crystal nanomembranes and strings \cite{saarinen2023laser,guo2019feedback}.  Anticipating and mitigating TINBA in these systems may therefore be a key step to operating them in the quantum regime.  In addition to increasing $Q$, a program combining multi-mode coherent \cite{saarinen2023laser} or measurement-based \cite{sommer2020multimode} feedback cooling, dual-probes at the ``magic detuning'' \cite{corbittAllOpticalTrapGramScale2007}, and, or, engineering of the effective mass and frequency spectrum \cite{galaNanomechanicalDesignStrategy2022}, may be advantageous.

\section{Acknowledgements}
This work is supported by NSF grant ECCS-1945832.  CMP acknowledges support from the ARCS Foundation, an Amherst College Fellowship, and a Grant in Aid of Research from Sigma Xi.  ARA acknowledges support from a CNRS-UArizona iGlobes fellowship.
Finally, the reactive ion etcher used for this study was funded by an NSF MRI grant, ECCS-1725571.
\color{black}

\bibliography{ref}
\end{document}



\title{Supporting information for: Thermal intermodulation back-action in a high-cooperativity optomechanical system}

\author{Christian M. Pluchar}
\author{Aman R. Agrawal}
\author{Dalziel J. Wilson}
\affiliation{Wyant College of Optical Sciences, University of Arizona, Tucson, AZ 85721, USA}
\date{\today}
\maketitle

\tableofcontents

\section{Experimental setup}

\begin{figure} [t!]
\vspace{-3mm}
    \includegraphics[width=1\columnwidth]{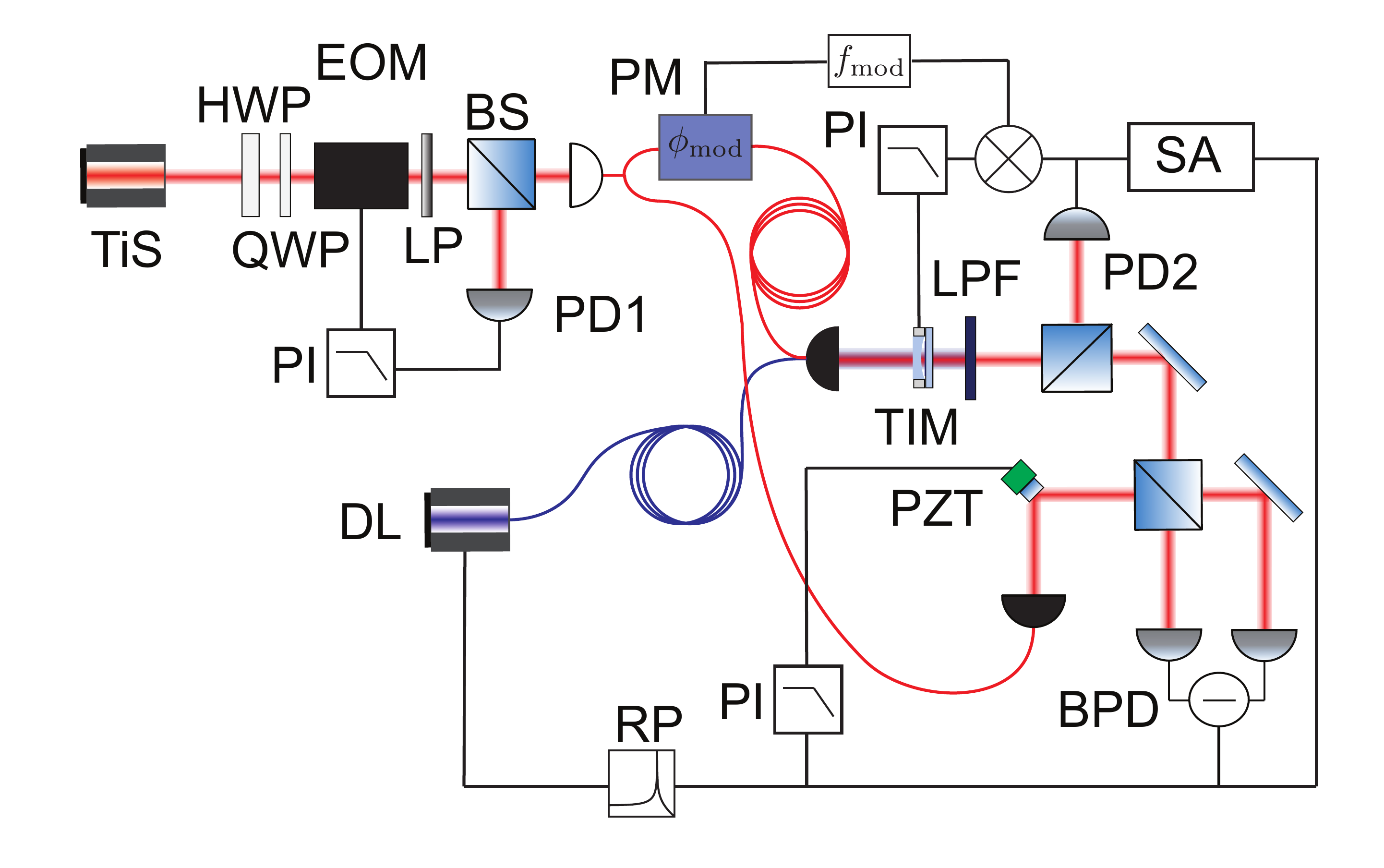}
    \raggedleft
    \caption{Experimental setup. HWP = half-wave plate, QWP = quarter-wave plate, EOM = electo-optic modulator, LP = linear polarizer, BS = beamsplitter, PI = proportional-integral controller, PM = phase modulator, PD1 = intensity suppression photodetector, PD2 = locking/intensity measurement photodetector, PZT = piezoelectric transducer, DL = diode laser, RP = Red Pitaya, LFP = long-pass optical filter, SA = spectrum analyzer, BPD = balanced photodetector.}
    \label{fig:detailedOpticalSetup}
\vspace{-3mm}
\end{figure}

Here, we describe our experimental setup in detail, including the apparatus for preparing the input optical field and characterizing the output field, and the feedback system for stabilizing laser intensity and laser-cavity detuning. \mbox{An overview is given in Fig. \ref{fig:detailedOpticalSetup}.}

\subsection{TIM cavity}

\subsubsection{Cavity assembly}
\begin{figure} [b!]
\vspace{-3mm}
\begin{center}
    \includegraphics[width=0.8\columnwidth]{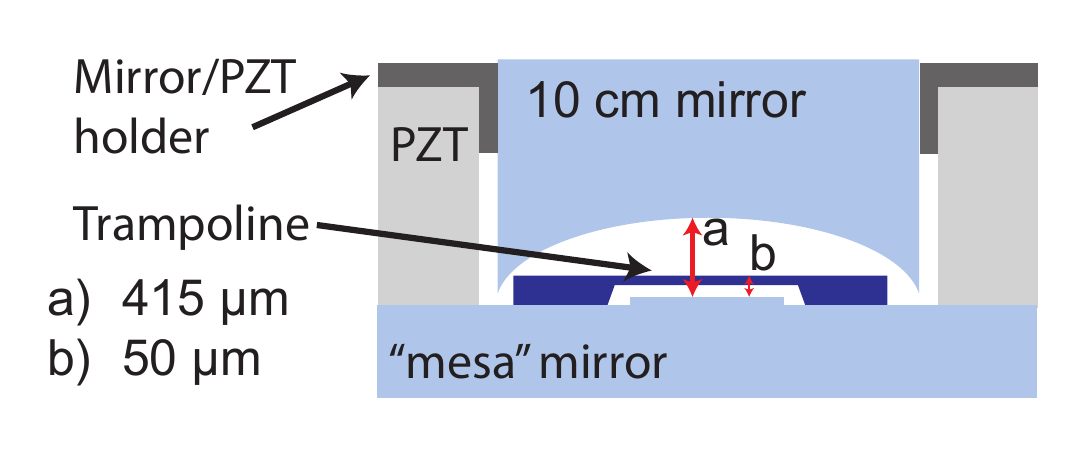}
    \caption{Sketch of the TIM cavity in cross-sction (not to scale). PZT = piezo-electric transducer. See text for details.}
    \label{fig:LabeledTIMSketch}
\end{center}
\vspace{-3mm}
\end{figure}
As shown in Fig. 2a, we place a high-$Q$ Si$_3$N$_4$ trampoline between two mirrors, forming a ``trampoline-in-the-middle'' (TIM) cavity optomechanical system \cite{jayich_dispersive_2008}.
The input mirror is concave with a 10 cm radius of curvature and the output mirror is planar.
The planar mirror is etched into a mesa-like structure 150 $\mu$m tall and 1 mm in diameter, allowing it to be recessed inside the etched portion of the 200-$\mu$m-thick Si device chip. The trampoline is thus positioned $50\,\mu\t{m}$ from the mirror coating, reducing diffraction loss due to wavefront curvature mismatch and in principle increasing the optomechanical coupling, via the ``membrane-at-the-edge" effect \cite{dumontFlexureTunedMembraneattheEdgeOptomechanical2019,wilson_cavity_nodate}. For the experiment in the main text, a nominal cavity length of $415\,\mu\t{m}$ was used (inferred from free spectral range measurements), employing a piezoelectric disk as a spacer (secured to the mesa mirror chip using two dabs of UV epoxy). 
The inferred mode radius at the trampoline position is roughly $40$ $\mu$m, implying less than 10 ppm clipping loss from the 200-$\mu$m-wide trampoline pad. The assembly is secured in compression with a leaf spring, forming a quasimonolithic cavity robust to~vibration. 


\subsubsection{Trampoline fabrication}
Following the trampoline fabrication process in \cite{reinhardt_ultralow-noise_2016, norte_mechanical_2016}, we begin our fabrication by coating a 1.5 $\mu$m thick photoresist (S1813) on a double-sided 100 nm thick Si$_3$N$_4$-on-Si wafer. The resist on one side of the wafer is patterned in the shape of a trampoline, and the other side is patterned with a square window using a standard backside alignment process in the photolithography system (MLA-150). After developing the exposed area, patterns on both sides of the wafer are transferred to the Si$_3$N$_4$ layer using a fluorine-based (Ar+SF$_6$) reactive ion etch. After removing the remaining resist, chips are cleaned using a 5-second dip in hydrofluoric acid (HF) followed by DI water and Isopropanol (IPA) rinse. The chips are then mounted on a custom Teflon holder to hold them in vertical orientation during the wet etch. Then the assembly is placed in a potassium hydroxide (KOH) bath at 80 °C for 21 hours to wet etch the Si in the patterned region and subsequently release the trampoline membrane on the other side of the chip. Finally, the released membrane is dried using a gradual dilution process, including iteratively replacing KOH with DI water followed by a 10-minute HF dip, IPA, and methanol rinse.

\begin{figure}[b]
\vspace{-10mm}
    \centering
    \includegraphics[width=0.85\columnwidth]{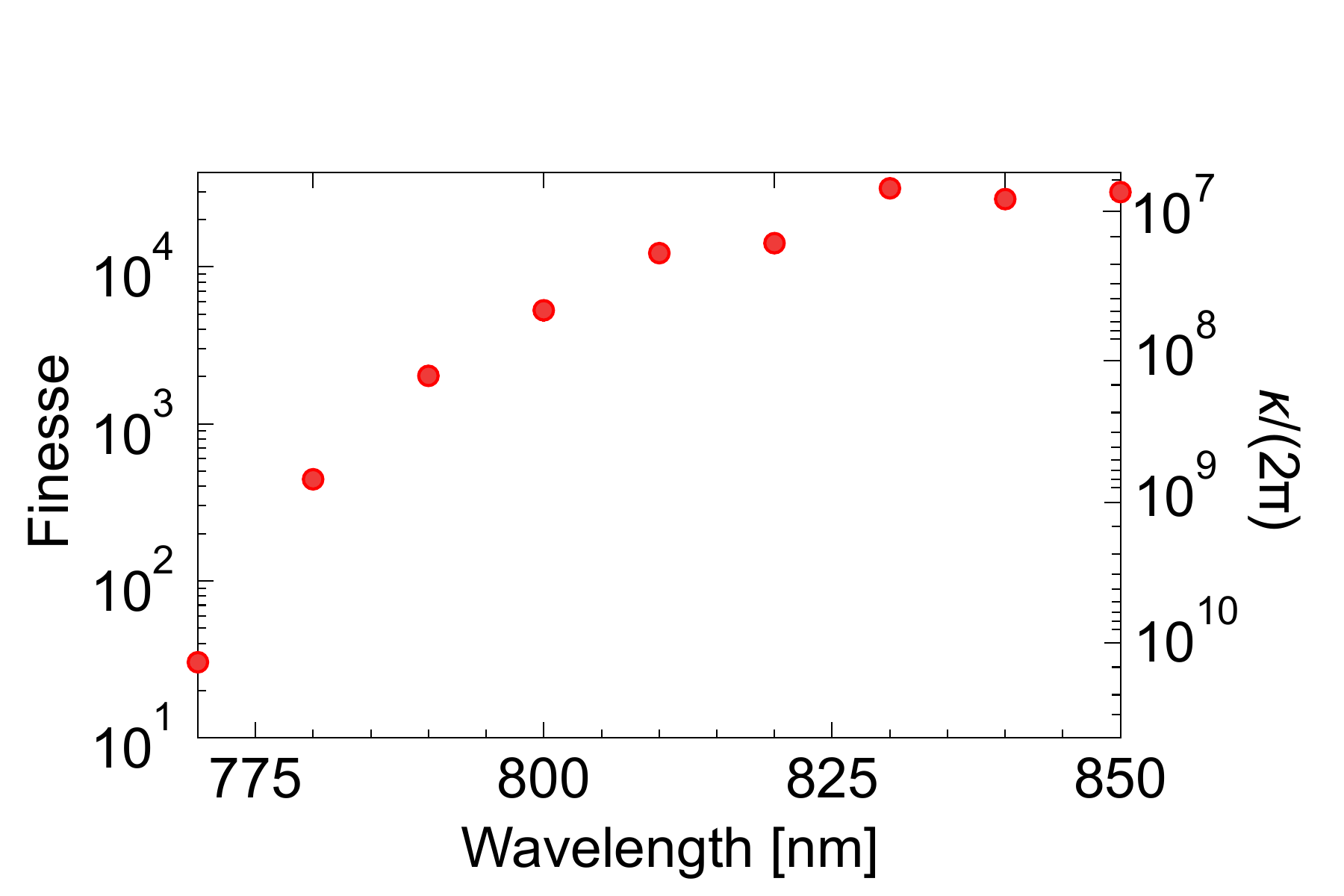}
    \caption{Measured finesse versus wavelength $\lambda$ for mirrors used in the experiment, without the trampoline inserted. The corresponding cavity linewidth for the TIM cavity length (415 $\mu$m) is shown on the right axis. Experiments in the main text were performed at $\lambda = 786$ nm.}
    \label{fig:emptyCavityCoatingCurve}
\vspace{-3mm}
\end{figure}

\subsubsection{Mirror coatings}
 
Measurements of the bare cavity finesse versus laser wavelength are shown in Fig. \ref{fig:emptyCavityCoatingCurve}.  The mirror coatings are nominally identical and support a cavity finesse of $\mathcal{F}\approx 3\times 10^4$ near the coating center wavelength of $\lambda\approx850$ nm.  As discussed in the main text, we deliberately operate at $\lambda = 786$ nm, corresponding to a finesse of $\mathcal{F}\approx 550$, to reduce thermal nonlinearity.

\subsubsection{Tuning the optomechanical coupling}

To tune the optomechanical coupling $G$, we rely on the fact that for an asymmetric membrane-in-the-middle system, the membrane's position on the intracavity standing wave depends on cavity mode order \cite{dumontFlexureTunedMembraneattheEdgeOptomechanical2019, nielsen2017multimode, wilson_cavity_nodate}.  We thus tune $G$ by coarsely tuning the laser wavelength and compensating with the piezoelectric spacer (which changes the gap between the trampoline and the curved mirror).

\subsection{Optical circuit}

\subsubsection{Titanium-Sapphire laser}

The laser used for our experiment is a continuous wave Titanium-Sapphire laser (MSquared Solstis, pumped by 10 W Laser Quantum Finesse).  The laser wavelength tuning range, 730 - 960 nm, is critical for coarsely tuning the cavity finesse (Fig. \ref{fig:emptyCavityCoatingCurve}).  At Fourier frequencies near the fundamental trampoline resonance ($\sim 40$ kHz), laser frequency noise, at the level 10 Hz/$\sqrt{\t{Hz}}$ (see Fig. \ref{fig:frequencyNoise}), was found to be negligible relative to thermal noise. 

\begin{figure}[b]
\vspace{-3mm}
    \centering
    \includegraphics[width=0.9\columnwidth]{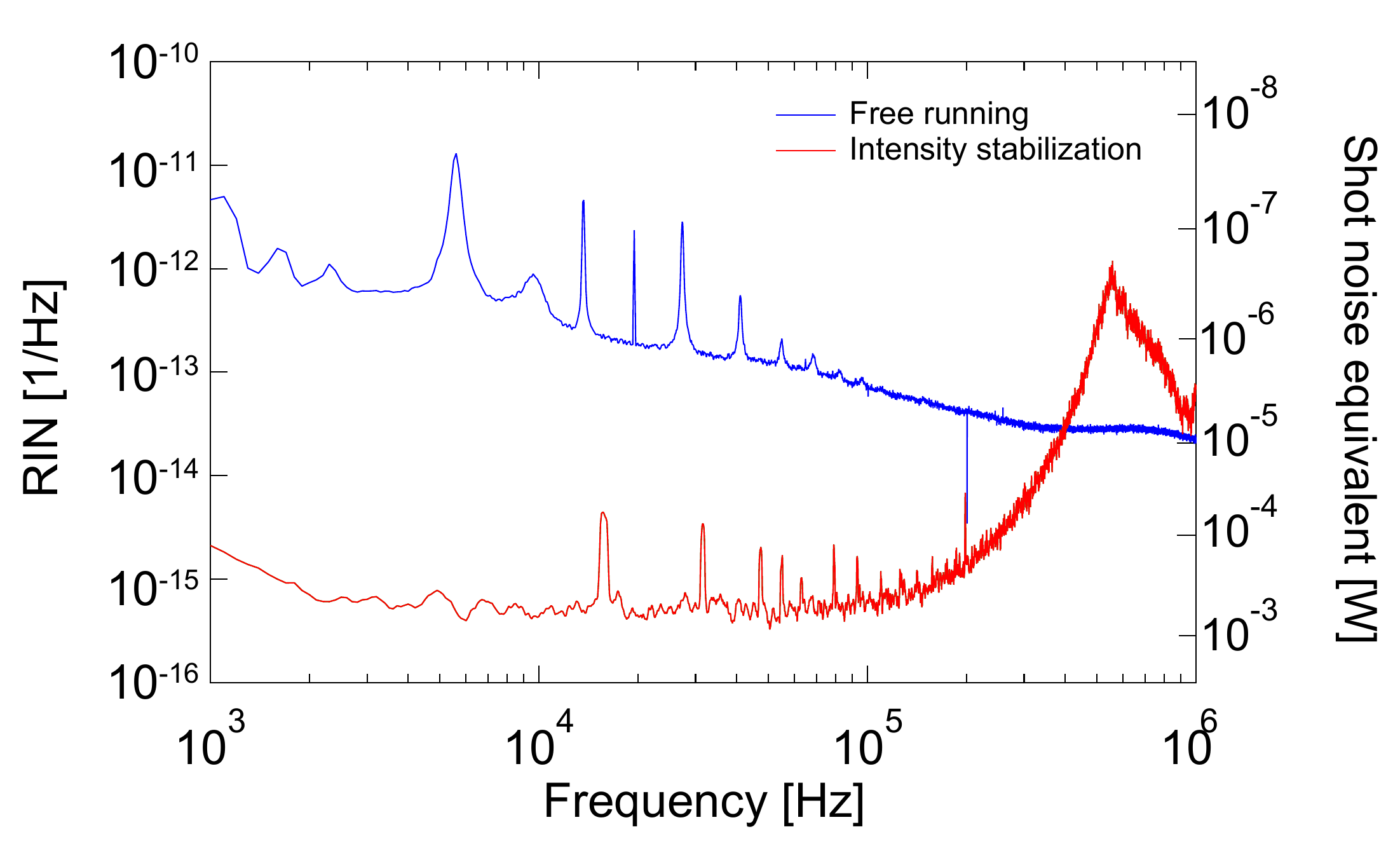}
    \caption{Free running and feedback-stabilized laser intensity noise in our experiment.  Shot-noise-equivalent power is shown on the right axis. In Figs. 3-4 of the main text, the stabilized intensity is far less than TIN.}
    \label{fig:noiseEatingPerformance}
\vspace{-3mm}
\end{figure}

\subsubsection{Intensity noise eater}

A key drawback of the Titanium-Sapphire (TiS) laser is intensity noise due to relaxation oscillation at 0.1 - 1 MHz frequencies.  The experiments shown in the main text require the laser intensity to be shot-noise-limited at $\sim 100$ kHz for powers as high as 1 mW, corresponding to $S_\t{RIN}\sim 10^{-15}\,\t{Hz}^{-1}$.  However, for our free-running TiS, we found $S_\t{RIN}\sim 10^{-13} \,\t{Hz}^{-1}$ at 100 kHz (due to relaxation oscillation and excess pump noise).  
To reduce this noise, we actively stabilized the laser intensity using a broadband electro-optic modulator (Thorlabs EO-AM-NR-C1) in conjunction with a low-noise, high dynamic range Si photodetector (Thorlabs PDA36A).  
Proportional-integral (PI) feedback was implemented using a high-speed analog controller (New Focus LB1005).  Gain settings were tuned to minimize noise near mechanical resonance.  As shown in Fig. \ref{fig:noiseEatingPerformance}, we were able to suppress intensity noise by 20 dB, to our target level $S_\t{RIN}\sim 10^{-15} \,\t{Hz}^{-1}$, from $1-200$ kHz.

\subsubsection{Input optics}

Downstream of the noise eater, the laser field is coupled into an optical fiber and split into two paths using a fiber beamsplitter: one to the cavity, and one to the local oscillator port of the homodyne receiver in cavity transmission.  Before coupling to the cavity, the light field is passed through a fiber electro-optic phase modulator (iXBlue NIR-MPX800-LN-10), for deriving the cavity lock error signal and calibrating optomechanical coupling.  Light is delivered into the cavity using a fiber-to-free-space coupler mounted on a 5-axis nanopositioning stage.  Mode-matching is refined by fine-tuning the distance between the fiber facet and the output coupling lens using a $z$-axis translation mount (Thorlabs SM1ZA).  For the experiments described in the main text, a typical mode-matching efficiency is $80\%$ (Sec \ref{sec:modeMatch}).

\subsubsection{Output optics and homodyne receiver}

 The light transmitted through the cavity (16\% of the mode matched power) is split into two unequal parts using a half wave plate and a polarizing beam splitter.  $\sim 5\%$ is sent to an avalanche photodetector (Thorlabs APD210) for intensity measurements and cavity locking. The remainder is directed to a balanced homodyne receiver employing a low noise balanced Si photodetector (Newport 1807).  The balanced homodyne receiver is used to record the phase of the output field---and therefore the displacement of the trampoline---as discussed in the main text.  The local oscillator (LO) phase is stabilized by applying PI feedback to a piezo-actuated mirror in the LO path, using a Newport LB1005 servo.

\subsubsection{Cavity lock}

For our experiment, the laser-cavity detuning is stabilized by feeding back to the cavity length via the piezo-electric mirror spacer (Fig. 2a).  A Pound-Drever-Hall error signal is generated by phase modulating the input field at $\sim 45$ MHz using the aforementioned fiber EOM.  The error signal is derived by demodulating the photocurrent of the avalanche photodiode in cavity transmission, at the same frequency. PI feedback is implemented with a Newport LB1005 servo. The unity gain frequency of the lock is less than 1 kHz and does not affect our thermal noise measurements.

\subsection{Feedback cooling}

\begin{figure}[b]
\vspace{-3mm}
    \centering
    \includegraphics[width=.95\columnwidth]{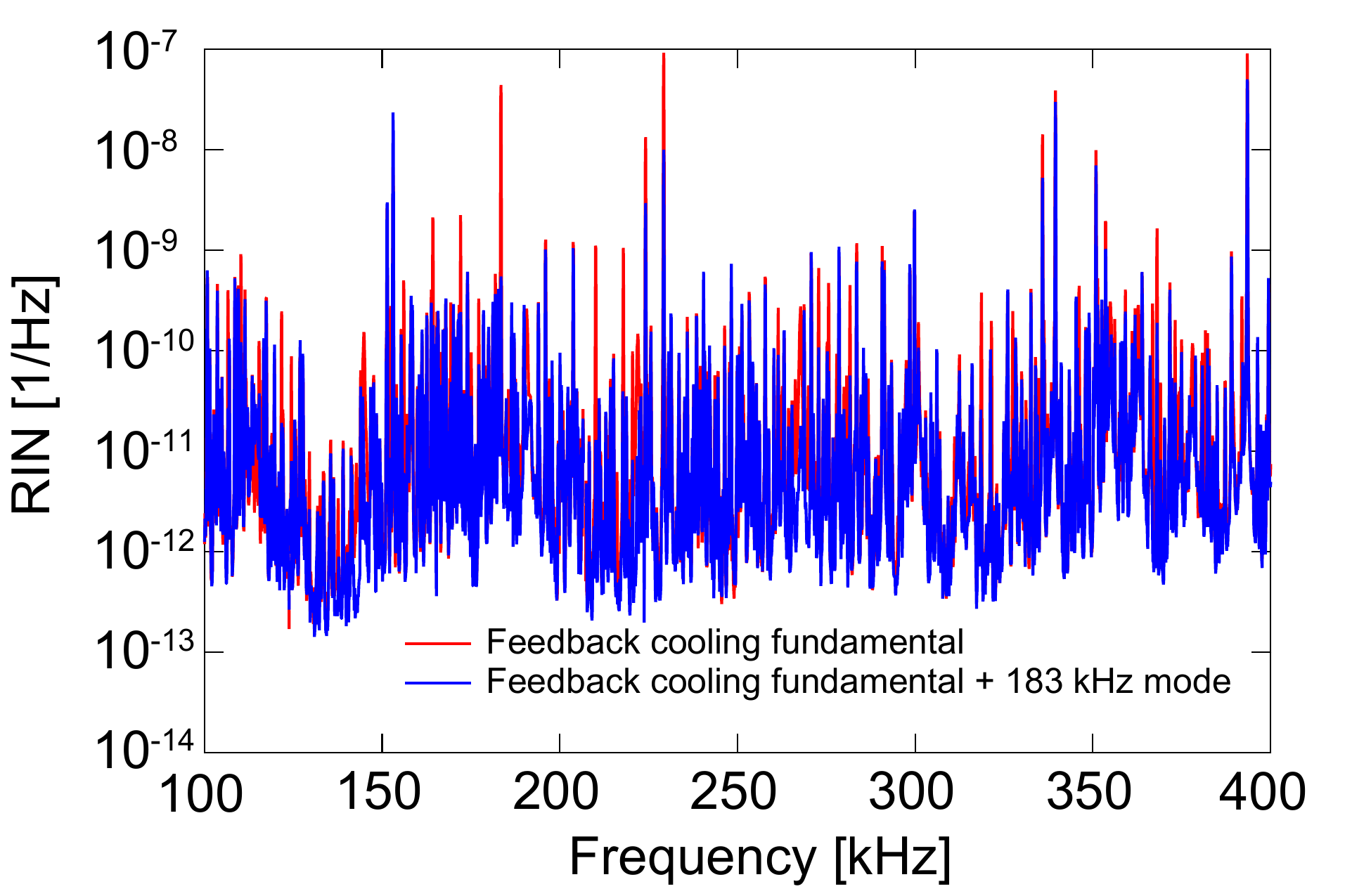}
    \caption{Comparison of TIN with active cooling the fundamental (41 kHz) mode (red) and fundamental plus second-order (183 kHz) mode (blue). Two-mode cooling suppresses the TIN peaks, and we measure a broadband noise reduction of 1.5 dB near the fundamental resonance.}
    \label{fig:feedbackCoolingModes}
\vspace{-3mm}
\end{figure}

For resonant measurements reported in the main text (Fig. 3-4), active feedback cooling was used to stabilize the fundamental trampoline mode \cite{pluchar_towards_2020}.  This had little effect on TIN near the fundamental resonance (which is produced by mixing of higher order modes) but prevented instabilities due to dynamical radiation pressure and photothermal backaction, and also reduced inhomogeneous broadening of the optical spring effect caused by thermal nonlinearity \cite{leijssen2017nonlinear}.  We also attempted multimode cooling of the first several trampoline modes but found that it had a marginal effect on TIN (see below).

To perform feedback cooling, we use the homodyne measurement as an error signal and apply feedback to the intensity of an auxiliary 650 nm diode laser (Thorlabs L650P007), through its current driver (Melles Griot 06DLD203).  The feedback laser is introduced via a fiber wavelength division multiplexer near the cavity input and is filtered from the output field using an optical long-pass filter (Thorlabs FEL0750).  Feedback control is implemented digitally via a Red Pitaya FPGA running an open-source software suite  (PyRPL \cite{neuhaus_pyrpl_2017}).  The use of an FPGA enables derivative feedback (approximated by a $90^\circ$ phase shift) to be applied over a narrow band of frequencies near one or several mechanical resonances, and was found to be crucial for our experiment.

\begin{figure*}[t!]
\vspace{-3mm}
    \centering
    \includegraphics[width=1.9\columnwidth]{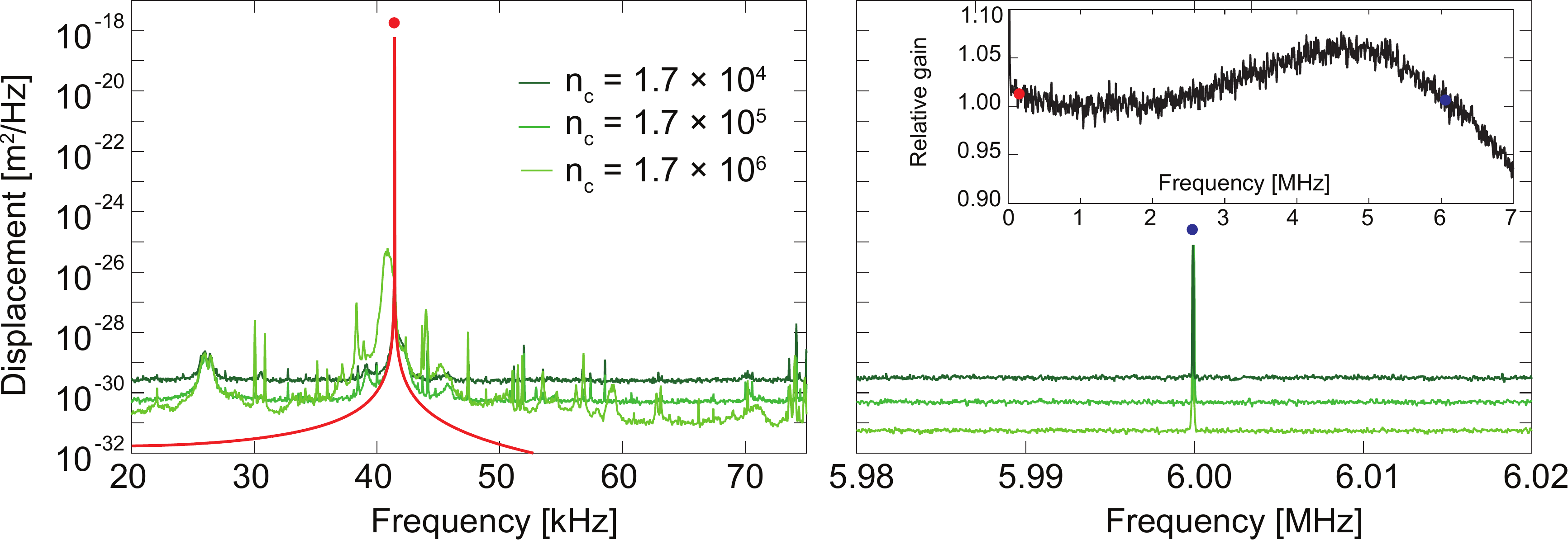}
    \caption{Calibration of vacuum optomechanical coupling and displacement (data from Fig. 4b). For low power measurements, we use the phase calibration tone at 6 MHz (right) to determine $g_0$ from the apparent displacement of the trampoline (left, see main text for details). We then calibrate all the remaining measurements using the phase calibration tone. The inset shows the detector response, with points indicating the mechanical frequency (red) and calibration tone frequency (blue).}
    \label{fig:vacuumOMC}
\vspace{-3mm}
\end{figure*}

As mentioned above, for the measurements in the main text, we cool only the fundamental mode.  It remains an open question whether TIN can be substantially reduced by multi-mode feedback cooling.  To explore this, as shown in Fig. \ref{fig:feedbackCoolingModes}, we cool the fundamental (41 kHz) and second order (183 kHz) trampoline mode. We observe a decrease in TIN peaks between 100-150 kHz and a smaller ($\sim$ 1.5 dB) reduction in the broadband noise.  This suggests that multimode cooling may indeed be effective at reducing TIN. However, in our case, reaching the required $\sim 30$ dB reduction for observing QBA seems likely to require cooling a prohibitively large number of modes, from a controller design standpoint.  In this regard, coherent feedback cooling using dynamical backaction remains a compelling alternative \cite{saarinen2023laser}.

\section{System characterization}
Here, we provide details on the measurements in Fig. 2 of the main text, which characterize our TIM system.

\subsubsection{Trampoline ringdowns and mode simulation}

Thermal noise models in Fig. 2b employ numerical simulations of the frequency and modeshape (and thereby the effective mass) of the first five transverse flexural modes of the trampoline.  These are obtained by finite element analysis using the COMSOL Multiphysics \cite{multiphysics1998introduction} Structural Mechanics module.

Ringdown measurements in Fig. 2c were performed by piezo-electrically exciting the membrane and monitoring the free energy decay using a real-time spectrum analyzer. (Data processing details may be found in the supporting information of \cite{pratt_nanoscale_2021}.)  Side-of-line cavity measurements were used for the mounted membrane and an independent Michelson interferometer was used for the unmounted membrane.  In both cases, the incident power was varied to ensure there was no photothermal damping or antidamping. For the cavity-based measurement, the 650 nm laser was used, with a cavity finesse of $\mathcal{F}\sim 1$, to minimize radiation pressure back action damping. We perform all ringdown measurements in high vacuum ($\sim 10^{-8}$ mbar), which we infer from monitoring the ion pump current.

\subsubsection{Calibrating detuning, optomechanical coupling, and displacement spectra}

Fig. 2d shows a measurement of the cavity transmission versus dynamically swept detuning, modulated by the thermal motion of the fundamental trampoline mode.  To calibrate the detuning (horizontal axis), an identical measurement (not shown) was made whilst the input field was phase modulated to create frequency sidebands at $\omega_\t{mod} = 2\pi\times 4\,\t{GHz}\approx 6.15 \kappa$.  As described in  the main text, fitting the thermally modulated steady-state response yields an estimate of the vacuum optomechanical coupling rate $g_0\approx 2\pi\times 1$ kHz.  We note however that this estimate is sensitive to the sweep rate (affecting dynamic backaction) and by stochastic fluctuations of the thermal displacement \cite{gieseler2014dynamic}, and is not reliable.

A separate estimate of the vacuum optomechanical coupling rate was made using the conventional frequency-modulation method \cite{gorodetksy_determination_2010}, and shown in Fig. \ref{fig:vacuumOMC}.  In this approach, the mean laser-cavity detuning is fixed on resonance while the laser is phase modulated with a calibrated $\beta$ at a frequency $ \omega_\t{mod} = 2 \pi \times 6$ MHz $\ll\kappa$. (We verify that the detector has a constant response over this bandwidth by examining its response to shot noise, as shown in the inset of Fig. \ref{fig:vacuumOMC}).  The resulting coherent tone in the balanced homodyne measurement has a spectral power (area of the peak in the power spectral density) of $(\beta\omega_\t{mod})^2/2$ in detuning units, and is used to calibrate $g_0$ by assigning the thermal noise peak a spectral power of $2g_0^2 k_B T_\t{eff}/(\hbar\omega_\t{m})$ in detuning units, where $T_\t{eff}$ is the effective mode temperature.  We perform the measurement using a weak resonant probe ($n_{\t{cav}} \approx 1400)$ in order to minimize dynamical backaction and assume that $T_\t{eff} \approx 300 $K.  This measurement yields $g_0 = 2\pi\times 1.5$ kHz used in our TIN and TINBA models. 

For displacement spectra in Figs. 2a, 3, and 4, we retain the phase modulation tone as a calibration reference. 
We do this by assigning the area of the tone an effective displacement variance of $(\beta\omega_\t{mod}/G)^2/2$.




\subsubsection{Input power and intracavity photon number} \label{sec:modeMatch}
In the main text, we relate input power $P_\t{in}$ and intracavity photon number $n_\t{c}$ using the formula
\begin{equation}
    n_\t{c}(\nu=0) = \frac{4\eta P_\t{in}}{\hbar\omega_0\kappa}
\end{equation}
where $n_\t{c}(\nu=0)$ is the value on resonance and $\eta\le 1$ is a factor related to cavity symmetry and mode-matching.

Optical spring measurements, as shown in Fig. 2e, are used to determine $n_\t{c}(\nu = 0)$.  For these measurements, we lock the cavity at different detunings, measure the mechanical resonance frequency shift $\Delta\omega_\t{opt}$ in the displacement spectrum, and compare the result to the following model \cite{aspelmeyer_cavity_2014} (valid in the bad cavity limit)
\begin{equation}
    \Delta \omega_\t{opt} = \frac{4  \nu g_0^2}{\kappa \left(1 + \nu ^2\right)^2} n_\t{c}(\nu=0).
\end{equation}
From these measurements we infer that $n_c = 1.7\times 10^6\times P_\t{in}/\t{mW}$, corresponding to $\eta = 0.40$.  For a symmetric cavity with an ideal value of $\eta = 0.50$, this implies a mode matching efficiency of approximately $80\%$.


\section{Data analysis}

In this section, we provide details on the intensity and phase noise spectra, and their models, in Figs. 3-4.
\subsection{Intensity noise}

\begin{figure}[t]
\vspace{-3mm}
    \centering
    \includegraphics[width=.95\columnwidth]{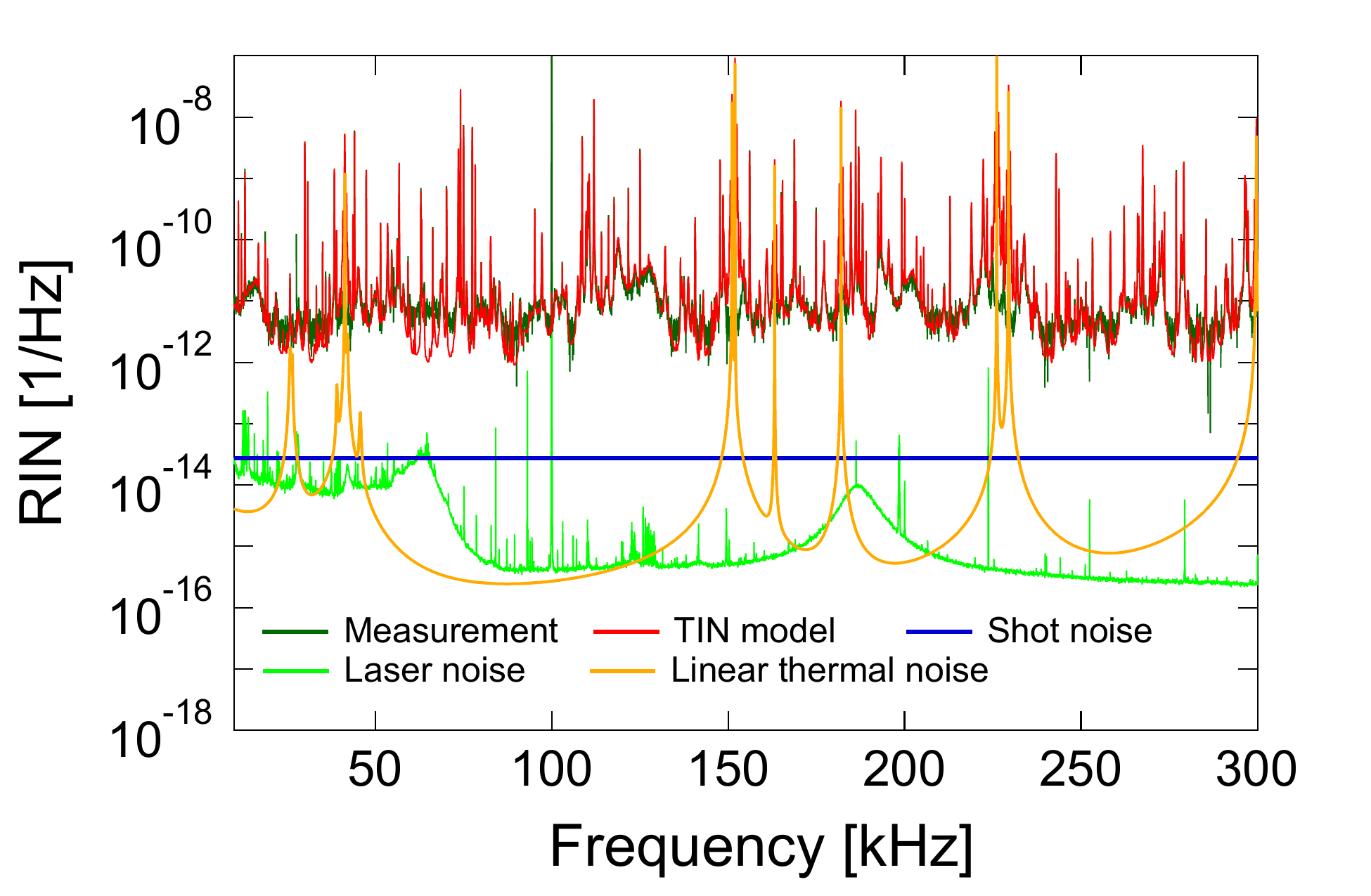}
    \caption{Full model of the intensity noise in Fig. 3b, including contribution due to linearly transduced thermal noise.}
    \label{fig:fullAmplitudeQuadModel}
\vspace{-3mm}
\end{figure}

In Fig. 3b we show measurements and models of the intensity noise in our TIM cavity, expressed in relative intensity noise (RIN) units.  A more detailed noise budget is shown in Fig. \ref{fig:fullAmplitudeQuadModel}, using the model:

\begin{subequations}\label{eq:SRIN}\begin{align}
    S_\t{RIN} [\omega] &= S_\t{RIN}^\t{shot}[\omega]+ S_{\t{RIN}}^\t{TIN}[\omega]+S^\t{th}_\t{RIN}[\omega] + S_\t{RIN}^\t{cl}[\omega]\\
     &\approx S_\t{RIN}^{\t{shot}}[\omega]+ S_{\t{RIN}}^{\t{TIN}}[\omega].
\end{align}\end{subequations}
Here $S_\t{RIN}^\t{shot}[\omega]$ and $S_\t{RIN}^\t{TIN}[\omega]$ are shot noise and TIN as described by Eq. 6 and 15, respectively; $S_\t{RIN}^\t{cl}[\omega]$ is classical laser intensity noise (e.g. relaxation oscillation noise), and $S_\t{RIN}^\t{th}[\omega]$ is residual linearly-transduced thermal noise.  In practice we find that the latter two terms are negligible at the frequencies of interest (1 - 200 kHz), due to our intensity noise eater and relatively large thermal nonlinearity.  This yields the approximation in Eq. \ref{eq:SRIN}b.


\subsubsection{Shot noise}

\begin{figure}[t]
       
    \centering
    \includegraphics[width=.95\columnwidth]{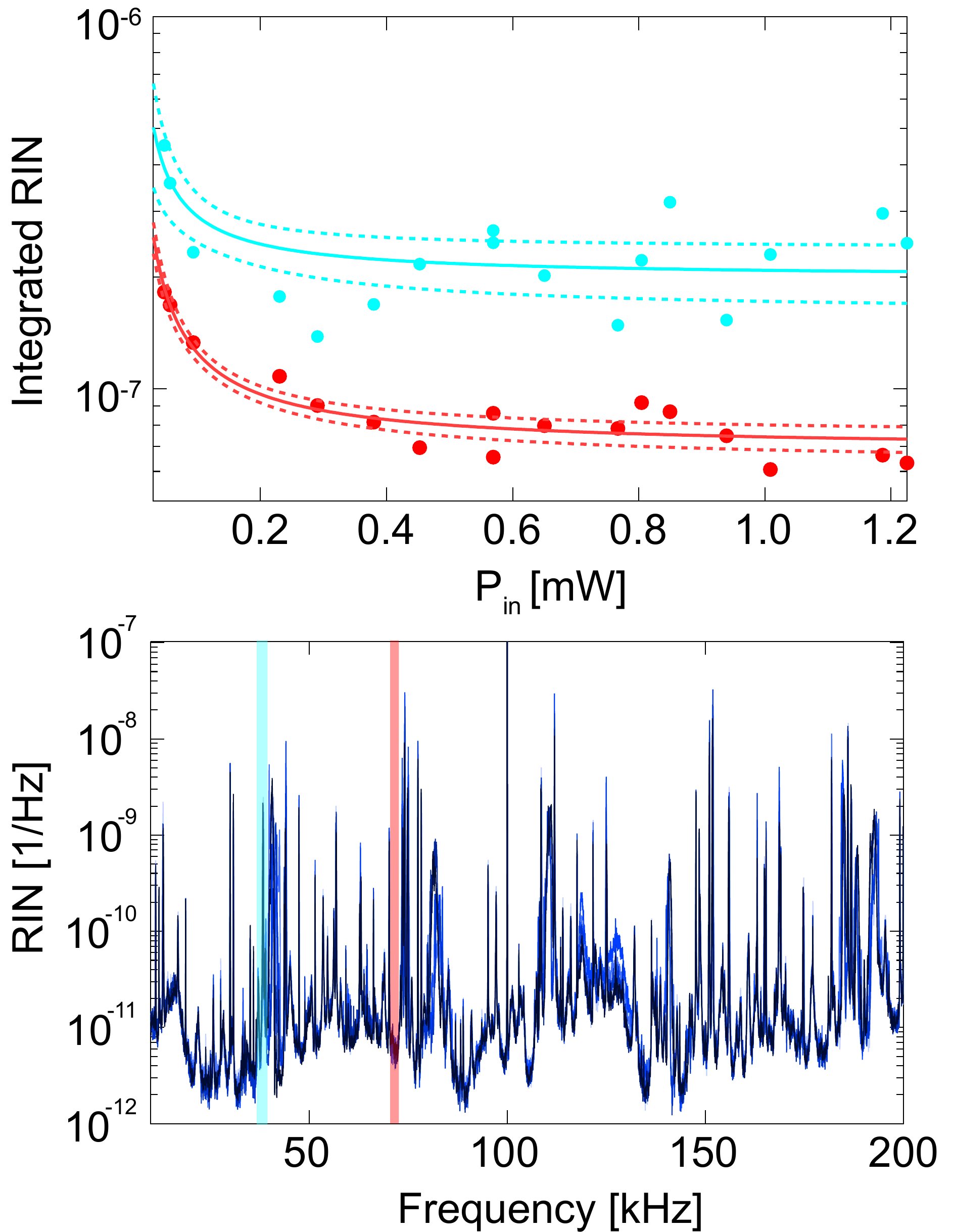}
    \caption{Integrated RIN versus in input power (top) for two frequency bands (bottom).  Solid lines are fits to the model $A P_{\t{in}}^{-1} + B$, where $A$ and $B$ parameterize shot noise and TIN, respectively.  
    Dashed lines correspond to 95\% fit confidence intervals. Variation in the noise level at large powers contributes to our uncertainty in the TINBA  model in Fig 4d.}
 \label{fig:RINVsPower}
\vspace{-3mm}
\end{figure}

The relative contributions of shot noise and TIN are discerned by varying input power $P_\t{in}$ (or equivalently, intracavity photon number $n_\t{c}$) and monitoring changes in $S_\t{RIN}[\omega]$. When $S_\t{RIN}[\omega]\propto 1/P_\t{in}$, it is assumed to be dominated by shot noise and when $S_\t{RIN}[\omega]$ is constant, it is assumed to be dominated by TIN. Plots of RIN versus power for the data in Fig. 3-4 are shown in Fig. \ref{fig:RINVsPower}a, integrated over two frequency bands as highlighted \ref{fig:RINVsPower}b.  At large powers, TIN appears to dominate shot noise by more than an order of magnitude.

\subsubsection{TIN}

As discussed in the main text, we confirm the observation of TIN by comparing it to Eq. 13, which predicts TIN based on the multimode thermomechanical noise spectrum (Eq. 14). The spectrum of the trampoline is difficult to model analytically, as it contains many modes with different $Q$ factors and effective masses. Instead, for each RIN measurement, we simultaneously record the phase noise of the output field using the homodyne detector.  This provides a measure of the apparent displacement $S_y[\omega] \approx S_x[\omega] + S_x^\t{imp}[\omega]$.  We then use $S_y[\omega] -S_x^{\t{imp}}[\omega]$ as the input to Eq. 13.

\subsubsection{Extraneous classical intensity noise}

Classical laser intensity noise $S_\t{RIN}^\t{cl}[\omega]$ and residual linearly-transduced thermal motion $S_\t{RIN}^\t{th} [\omega]$ contribute marginally to the total noise budget in Fig. \ref{fig:fullAmplitudeQuadModel}.  Classical laser intensity noise is found to be negligible at the powers used.  This is determined by direct detection of the light field incident upon the cavity (gray spectrum in Fig. \ref{fig:fullAmplitudeQuadModel}).  Linearly transduced thermal noise arises from small nonzero detunings from resonance (the second term in Eq. 12).  This noise predominates near mechanical resonance frequencies $\omega_n$---unlike TIN, which occurs at frequencies $\omega_n \pm \omega_n'$.  By fitting higher order residual thermal noise noise peaks $S_\t{RIN}[\omega_n]$ to a multimode thermal noise models (orange curve in Fig. \ref{fig:fullAmplitudeQuadModel}), we estimate that $S_\t{RIN}^\t{TIN}[\omega] / S_\t{RIN}^\t{th} [\omega] \gg 30$ dB at frequencies away from mechanical resonances, including those analyzed for Fig. 4 of the main text. 

\begin{figure}[b!]
\vspace{-3mm}
    \centering
    \includegraphics[width=.95\columnwidth]{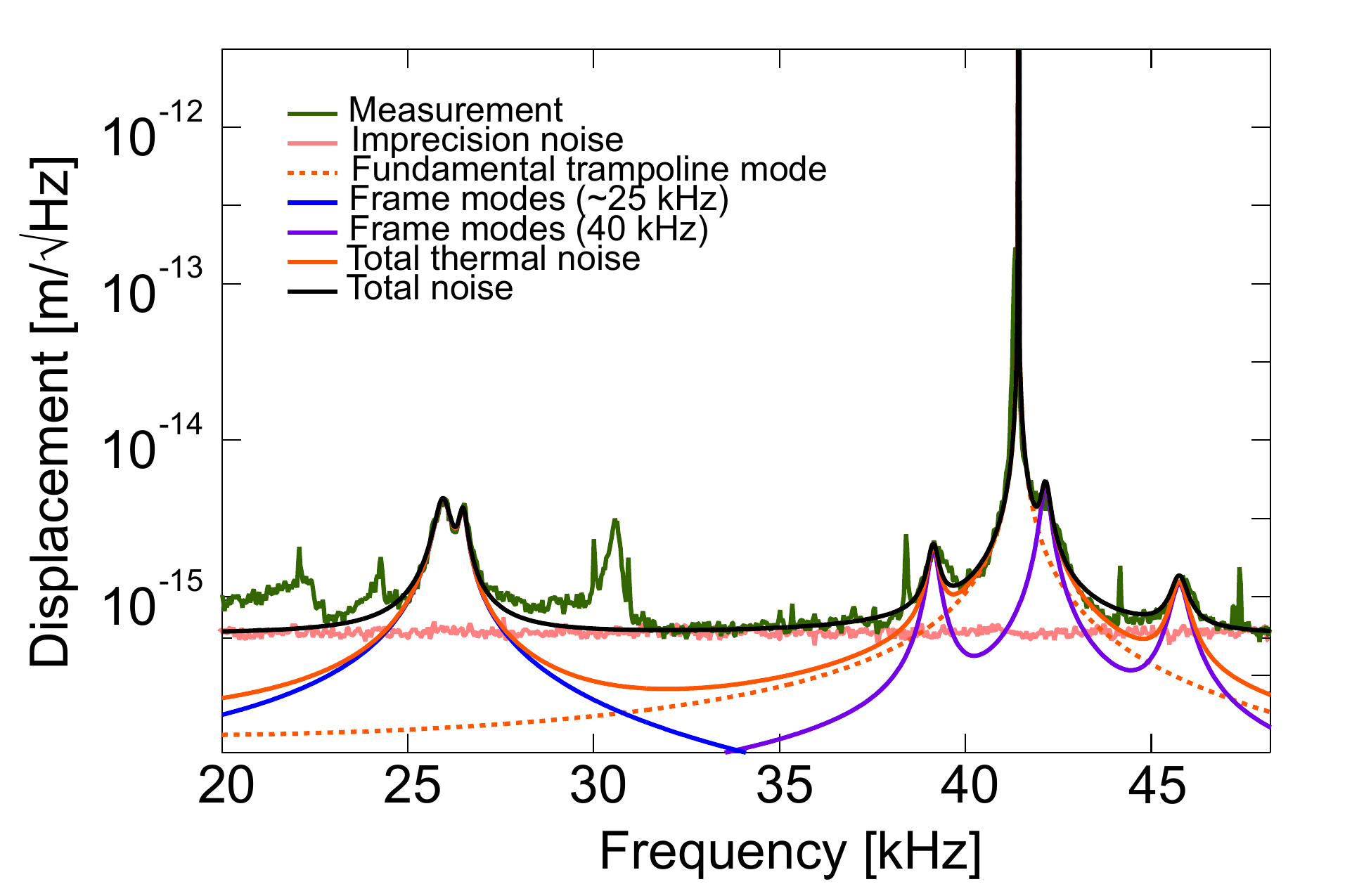}
    \caption{Apparent displacement noise near the fundamental trampoline resonance. In addition to shot noise and trampoline thermal motion, we observe extraneous noise peaks that are believed to be due to thermal motion of the Si device chip (purple).  For this measurement, $P_{\t{in}} = 30 \mu$W ($n_c = 5 \times 10^4$), so the back-action and photothermal effects are negligible.}
\label{fig:noiseBudgetWithExtraThermalNoise}
\vspace{-3mm}
\end{figure}

\subsection{Phase (displacement) noise}

In Fig. 3a we show measurements and models of the phase noise of the TIM output field, expressed in apparent displacement noise (RIN) units.  A more detailed noise budget is shown in Fig. \ref{fig:fullAmplitudeQuadModel} (for an independent measurement with smaller input power), using the model:
\begin{subequations} \label{eq:totalNoiseBudget}\begin{align}
    S_{y}[\omega] & = S_x^\t{imp,shot}[\omega] + S_x^\t{imp,TIN}[\omega]  + S_x^\t{imp,f}[\omega]\\ &+  S_x^\t{th}[\omega] + S_x^\t{QBA}[\omega] + S_{x}^{\t{TIN}}[\omega] + S_x^{\t{CBA}}[\omega] \\
    & \approx S_{x}^{\t{imp}}[\omega] + S_{x}^{\t{th}}[\omega] + S_{x}^{\t{TIN}}[\omega].
\end{align}\end{subequations}
Here $S_x^\t{imp}[\omega]$ the phase readout noise with components due to shot noise $S_x^\t{imp,shot}[\omega]$, phase TIN $S_x^\t{imp,TIN}[\omega]$, and laser frequency noise $S_x^\t{imp,f}[\omega]$, respectively; and $S_x^\t{th}[\omega]$, $S_x^\t{QBA}[\omega]$, $S_x^\t{TIN}[\omega]$, and $S_x^\t{CBA}[\omega]$ are thermal motion, QBA-driven motion, TINBA-driven motion, and motion driven by other forms of classical backaction such as photothermal heating, respectively.  Among the different forms of readout and physical noise, shot noise, thermal noise, and TINBA-driven motion dominate our measurements, yielding the approximation in \ref{eq:totalNoiseBudget}c. 


\subsubsection{Shot noise}

As shown in Fig. \ref{fig:noiseBudgetWithExtraThermalNoise}, shot noise dominates the imprecision of our homodyne readout.  We determine this by blocking the signal arm of our homodyne interferometer, leaving only the shot noise of the local oscillator.  For our measurements, the LO power is $P_\t{LO} = 4\,\t{mW}\gg P_\t{sig}$, so its shot noise is a good approximation of $S_x^\t{imp,shot}[\omega]$.  

\subsubsection{Thermal noise}
\label{sec:thermalNoise}
Thermal noise models in Figs. 2-4 were generated by fitting each peak to a model including optical damping \cite{aspelmeyer_cavity_2014} and structural internal damping \cite{saulson_thermal_1990}
\begin{equation}
S_x[\omega] = \left|\chi_\t{eff}[\omega]\right|^2 4 k_{B} T m\gamma_\t{m} \omega_\t{m}/\omega,
\end{equation}
where 
\begin{equation}
\chi_\t{eff}[\omega] = \frac{1/m}{((\omega_\t{m}+\Delta\omega_\t{opt})^2-\omega^2)+i\omega(\gamma_\t{m}+\gamma_\t{opt})}
\end{equation}
is the effective mechanical susceptibility including the optical spring shift $\Delta\omega_\t{opt}$ and damping $\gamma_\t{opt}$, and $T = 298$ K is the bath temperature (assumed to be room temperature). We do this at low powers where the effect of TINBA or other forms of heating is negligible. For higher order modes, we assume that $\Delta\omega_\t{opt}$ and $\gamma_\t{opt}$ are small.  Otherwise $\gamma_\t{m}$ and $m$ are left as free parameters. When fitting, the points around mechanical resonance are masked, since the resolution of our spectrum analyzer is greater than the mechanical linewidth.  As mentioned in the main text and highlight in Fig. 4a, We also fit several noise peaks around the fundamental resonance frequency, which we suspect arise due to coupling between the fundamental mode and the frame modes of the chip \cite{dejongMechanicalDissipationSubstratemode2022}.  The $Q$ of these modes is $Q \sim 10^3$, which is consistent with the intrinsic $Q$ of the Si chip material.  Their effective mass is also consistent with the physical chip mass, $\rho_\t{Si} \times$ (1 cm)$^2$ $\times$ 100 nm $\approx 2 \times 10^{-8}$ kg, where $\rho_\t{Si} \approx 2400$ kg/m$^3$ is the density of Si.   

After fitting individual peaks, we incoherently sum them to produce the multimode thermal noise models shown in the main text and in Fig. \ref{fig:noiseBudgetWithExtraThermalNoise}. 

\begin{figure}[t]
\vspace{-3mm}
    \centering
    \includegraphics[width=.95\columnwidth]{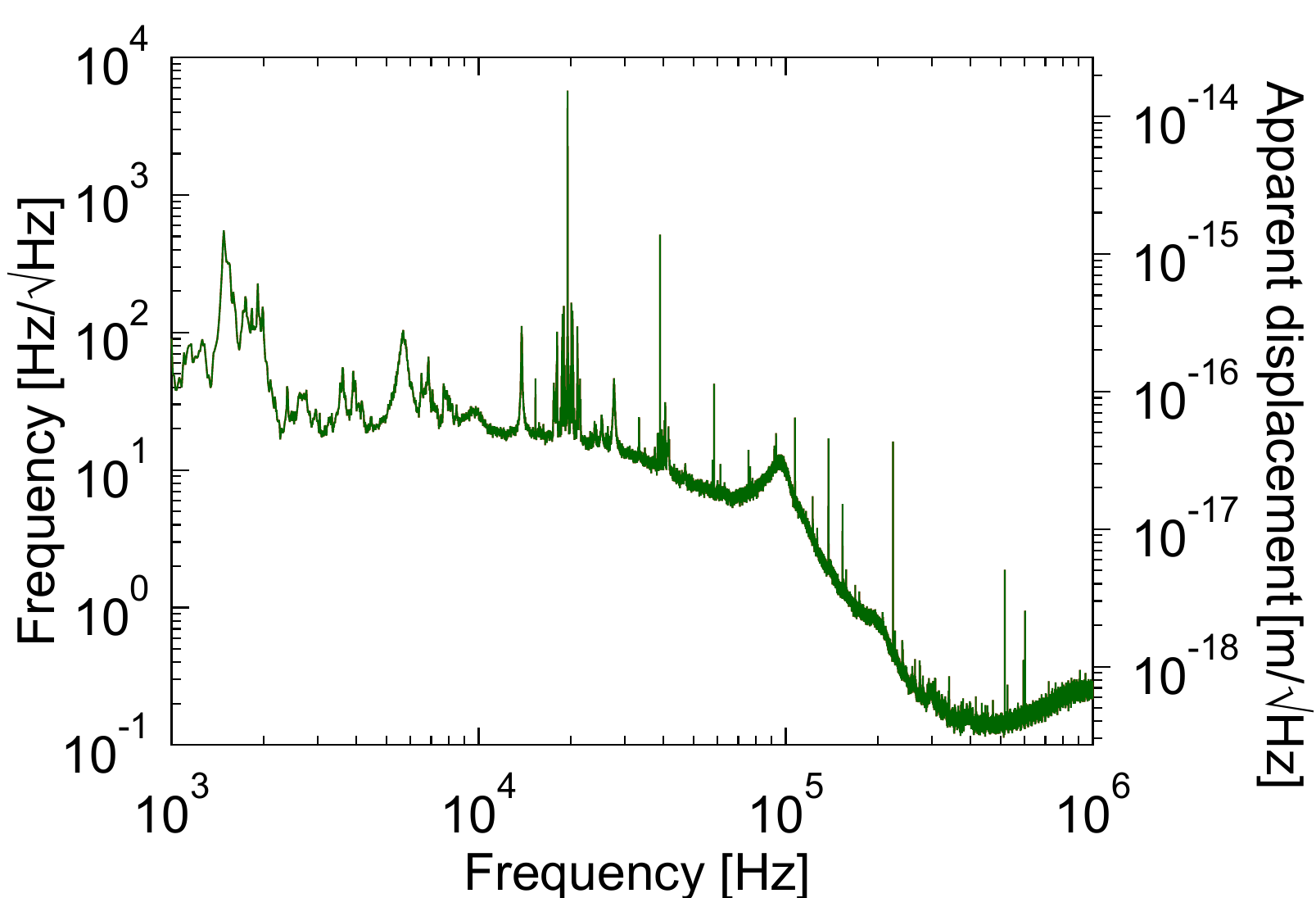}
    \caption{Laser frequency noise measured between 1 kHz (the reference cavity lock bandwidth) and 320 kHz (the reference cavity linewidth). The equivalent trampoline displacement noise is shown on the right axis.}
    \label{fig:frequencyNoise}
\vspace{-3mm}
\end{figure}

\subsubsection{Laser frequency noise}

Our short cavity is relatively immune to laser frequency noise $S_f[\omega]$, since the apparent diplacement noise it produces scales as $S_x^{\t{imp},f}[\omega] = S_f[\omega]/G^2 \propto  L^2 S_f[\omega]$, where $L$ is the cavity length.  To confirm this, as shown in Fig. \ref{fig:frequencyNoise}, we measured the frequency noise of our Titanium Sapphire laser by passing it through a 10 cm long monolithic reference cavity and monitoring the transmitted power on the side of the line \cite{kippenbergPhaseNoiseMeasurement2013}. We infer $\sqrt{S_f[\omega]} \approx$ 10 Hz/$\sqrt{\t{Hz}}$ in the $10-100$ kHz frequency range, corresponding to an apparent displacement of  $S_x^{\t{imp}, f} [\omega_m] \approx 2 \times 10^{-17}$ m/$\sqrt{\t{Hz}}$.  This noise level contributes negligibly to the measurement in \ref{fig:noiseBudgetWithExtraThermalNoise}.

\subsubsection{TINBA}
The TINBA model in Fig. 4d is generated from Eq. 16c in the main text and with measured parameters $G, n_c,$ and $S_{\t{TIN}}^{\t{RIN}}[\omega]$. The uncertainty in the TINBA magnitude comes from variation in $S_{\t{TIN}}^{\t{RIN}}[\omega]$, which we suspect varies due to slight changes in detuning and the stochastic variation in the thermal amplitude of the mechanical modes, as we measure for less than the mechanical coherence time of many of the modes. The mean level (solid line) is set by taking the magnitude of the largest power measurement over the frequency band and scaling by $P_{\t{in}}^2.$

\subsection{Correlation measurement}

\begin{figure}[b]
\centering
    \vspace{-3mm}
    \includegraphics[width=0.9\columnwidth]{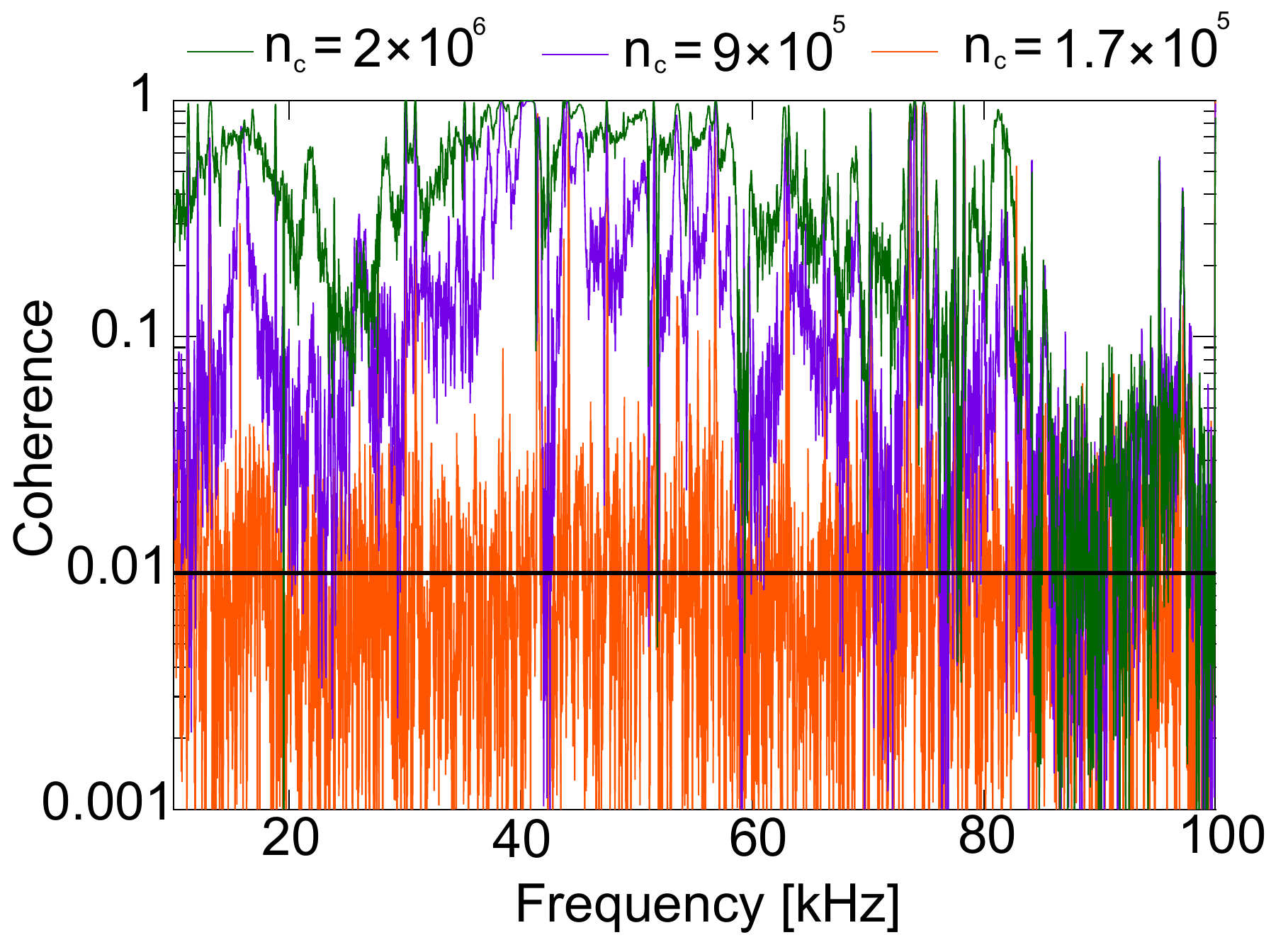}
    \caption{Displacement-RIN coherence for different intracavity photon numbers. Black line: bias of the estimated cohe
rence for the number of measurements we collect, $N = 100$.}
\label{fig:coherenceVsPower}
\end{figure}

As discussed in the main text, we measure the correlation between phase and intensity noise measurements to further confirm the contribution of TINBA.  We compute the coherence, the normalized cross-spectrum:
\begin{equation}
    C_{ab}[\omega] = \frac{|S_{ab}[\omega]|^2}{S_{a}[\omega] S_{b}[\omega]}\in[0,1]
\end{equation}
where $a$ and $b$ are two signals with power spectra $S_{a,b}[\omega]$ and cross spectrum $S_{ab}[\omega]$.  This function is real-valued, so to capture the full dynamics, one must also examine the phase of the cross-spectrum $\t{Arg}[S_{ab}]$.  In our case $S_{a}[\omega] =  S_{y}[\omega]$ and $S_{b}[\omega] = S_{\t{RIN}}[\omega]$.  

The measurement in Fig. 4c is made at sufficiently large power that TINBA dominates $S_y[\omega]$ and TIN in $S_\t{RIN}[\omega]$ over a large range of frequencies near the fundamental mechanical resonance, corresponding to $C_{ab}[\omega] = 1$.  In  Fig. \ref{fig:coherenceVsPower} we overlay coherence measurements at two intermediate powers (parameterized by the $n_c$), to visualize the increase in correlation as a function of measurement strength. At lower powers and frequencies far from the mechanical resonance---where we expect the cross-correlation to be small---our measured coherence asymptotes to \textcolor{black}{the bias of the  estimate for $N = 100$ finite-time samples, Bias$[\hat{C}_{ab}[\omega]] = 1/N = .01$ \cite{carterEstimationMagnitudesquaredCoherence1973}.}

With regards to other processes contributing correlations between the phase and amplitude measurements \cite{borkje_observability_2010}:  we rule out correlations due to $S^\t{imp, RIN}$ due to its low contribution to the phase noise spectrum, different frequency dependence (TINBA is weighted by the mechanical susceptibility), and the $\pi$-phase shift of $\arg[C_{ab}]$ on mechanical resonance, indicative of the mechanical susceptibility. Contributions from thermal noise are present in both spectra, but only at frequencies very close to the mechanical resonance in the amplitude quadrature (Fig. \ref{fig:fullAmplitudeQuadModel}). Additionally, thermal noise will have a constant $C_{ab}$ as a function of power, unlike TINBA or another form of radiation pressure driving.

\color{black}
Finally, we note that the model overlayed onto the correlation measurement in Fig. 4b is \cite{verlotExperimentalDemonstrationQuantum2011, purdy_observation_2013} 
\begin{equation}
    C_{ab}[\omega] \approx \frac{S_x^\t{TIN}[\omega]} {S_{y}[\omega]}\frac{S_\t{RIN}^\t{TIN}[\omega]}{S_\t{RIN}[\omega]},
\end{equation}
where the first term represents the fraction of the mechanical motion due to TINBA, and the second is the percentage of the intensity noise due to TIN, which is a good approximation in the absence of other correlated noise processes.
\color{black}

\bibliography{SIRef}